\documentclass[12pt,epsfig]{article}
\usepackage[xdvi]{graphicx}

\newcommand{\beq}{\begin{equation}}
\newcommand{\eeq}{\end{equation}}
\newcommand{\bea}{\begin{eqnarray}}
\newcommand{\eea}{\end{eqnarray}}

\def\e2sig{e^{-2r\sigma}}

\makeatletter

\@addtoreset{equation}{section}
\makeatother

\begin{document}
\setlength{\baselineskip}{18pt}

\begin{titlepage}
\begin{flushright}
KOBE-TH-08-06
\end{flushright}

\vspace{1.0cm}
\begin{center}
{\Large\bf More on the Finiteness of Anomalous Magnetic \\ 
\vspace*{5mm}
Moment in the Gauge-Higgs Unification} 
\end{center}
\vspace{25mm}

\begin{center}
{\large
Yuki Adachi, 
%
C. S. Lim
%
%
and Nobuhito Maru$^*$
}
\end{center}
\vspace{1cm}
\centerline{{\it Department of Physics, Kobe University,
Kobe 657-8501, Japan}}

\centerline{{\it
$^*$Department of Physics, Chuo University, 
Tokyo 112-8551, Japan
}}
%
%
\vspace{2cm}
\centerline{\large\bf Abstract}
\vspace{0.5cm}
We discuss the anomalous magnetic moment of fermion 
in a realistic $SU(3)$ model of gauge-Higgs unification compactified on an orbifold $S^{1}/Z_{2}$ 
including $Z_{2}$-odd bulk mass for fermions. 
An operator analysis implies that the anomalous magnetic moment should be finite and predictable, 
even though higher dimensional gauge theories are non-renormalizable. 
Our main purpose is to clarify the cancellation mechanism of the UV divergences 
appearing in various types of Feynman diagrams. 
The cancellation of divergence turns out to take place 
between the contributions of the ``partners" in the Higgs-like mechanism 
present in the non-zero Kaluza-Klein modes to form massive gauge bosons. 
It is also argued that the cancellation may be attributed 
to the quantum mechanical SUSY hidden in the scenario of gauge-Higgs unification. 
\end{titlepage}

\section{Introduction}

Gauge-Higgs unification \cite{Manton, Fairlie, Hosotani} is one of the attractive scenarios 
solving the hierarchy problem without invoking supersymmetry. 
In this scenario, 
Higgs doublet in the Standard Model (SM) is identified with 
the extra spatial components of the higher dimensional gauge fields. 
Remarkable feature is that the quantum correction to Higgs mass is insensitive 
to the cutoff scale of the theory and calculable 
regardless of the non-renormalizability of higher dimensional gauge theory. 
The reason is that the Higgs mass term as a local operator is forbidden 
by the higher dimensional gauge invariance. 
The finite mass term is generated radiatively and expressed by the Wilson line phase as a non-local operator. 
This fact has opened up a new avenue to the solution of the hierarchy problem \cite{HIL}. 
Since then, much attention has been paid to the gauge-Higgs unification and 
many interesting works have been done from various points of view 
\cite{KLY}-\cite{LHC}. 

The finiteness of Higgs mass has been studied and verified in various models 
and types of compactification at one-loop level\footnote{For the case of gravity-gauge-Higgs unification, 
see \cite{HLM}} \cite{ABQ}-\cite{LMH} and even at two loop level \cite{MY, HMTY}. 
It is natural to ask whether any other finite physical observables exist in the gauge-Higgs unification. 
The naive guess is that such observables are in the gauge-Higgs sector of the theory if they ever exist. 
Two of the present authors (C.S.L. and N.M.) studied the structure of divergences for S and T parameters 
in the gauge-Higgs unification since such parameters are described 
by higher dimensional gauge invariant operators with respect to gauge and Higgs fields, 
and are expected to be finite by virtue of the higher dimensional gauge symmetry. 
The result is that both parameters are divergent (convergent) more than (in) five dimensions 
as expected from the naive power counting argument.  
However, a nontrivial prediction  we have found, specific to the gauge-Higgs unification, is 
that some linear combination of S and T parameters is finite even in six dimensions \cite{LM}. 

In our previous paper \cite{ALM} we have found a more striking fact:  
we have shown that the magnetic moment of fermion 
in the $(D+1)$ dimensional QED gauge-Higgs unification model 
compactified on $S^1$ becomes finite for an arbitrary space-time dimension, 
regardless of the nonrenormalizability of the theory. 
Actually, the reason is very simple. 
In four dimensional space-time, 
a dimension six gauge invariant local operator describes the magnetic moment: 
\bea
i \bar{\psi}_L \sigma^{\mu\nu} \psi_R F_{\mu\nu} \langle H \rangle. 
\label{MMM4}
\eea
However, when included into the scheme of gauge-Higgs unification, 
the Higgs doublet should be replaced by an extra space component of the higher dimensional gauge field $A_y$. 
Then the operator is forbidden by the higher dimensional gauge invariance, 
since $A_y$ transforms inhomogeneously under the gauge transformation. 
Then, to preserve the gauge symmetry, 
$A_y$ should be further replaced by gauge covariant derivative $D_y$, 
and the relevant gauge invariant operator becomes 
\bea
i \bar{\Psi} \Gamma^{MN} D_L \Gamma^L \Psi F_{MN}  
\label{MMMD}
\eea
where $L, M$ and $N$ denote $D+1$ dimensional Lorentz indices. 
The key observation of our argument is that the operator (\ref{MMMD}), 
when $D_L$ is replaced by $\langle D_L \rangle$ with the gauge field $A_L$ replaced by its VEV, 
vanishes because of the on-shell condition $i \langle D_L \rangle \Gamma^L \Psi = 0$. 
From this fact, we can expect that the magnetic moment is finite 
and have shown that it is indeed the case by explicit diagrammatical calculations \cite{ALM}. 
This is the specific prediction of the gauge-Higgs unification to be contrasted with 
the case of Randall-Sundrum model \cite{DHR} or the universal extra dimension scenario \cite{ADW, AD}, 
in which the magnetic moment of fermion diverges in the models with more than five space-time dimensions. 

Although this result was quite remarkable, the above model is too simple and not realistic. 
In particular, 
the gauge group $U(1)$ is too small to incorporate the standard model. 
In this paper, we study more about the cancellation mechanism of ultraviolet (UV) divergences 
in a realistic gauge-Higgs unification model. 
We consider $(D+1)$ dimensional $SU(3)$ gauge-Higgs unification model compactified 
on an orbifold $S^1/Z_2$ with a massive bulk fermion in a fundamental representation. 
The orbifolding is indispensable to obtain the SM Higgs $SU(2)_L$ doublet 
since Higgs originally behaves as an adjoint representation of the gauge group in the gauge-Higgs unification. 
We consider here a simple orbifold $S^1/Z_2$. 
In the case of $S^1/Z_2$, the bulk mass parameter of fermion must have odd $Z_2$ parity 
since the fermion bulk mass term connects fermions with different chiralities and opposite $Z_2$ parities. 
It is well known that the zero mode wave functions take an exponential profile along a compactified space coordinate 
and $D$-dimensional effective Yukawa couplings obtained by an overlap integral of zero mode wave functions 
are exponentially suppressed. 
In this way, we can freely obtain the light fermion masses, which are otherwise of ${\cal O}(M_W)$, 
by tuning the bulk mass parameter. 
One might worry if our argument for the finiteness in the previous paper 
still holds in the present orbifold model since the on-shell condition for the fermion is changed to 
$i \Gamma^M \langle D_M \rangle \Psi = M \varepsilon(y) \Psi$ 
($\varepsilon(y):$ the sign function of $y$, the extra space coordinate) and also the brane localized operator 
\bea
i \bar{\psi}_L \Gamma^{\mu\nu} A_y \Gamma^y \psi_R F_{\mu\nu} 
\label{braneAMM}
\eea
seems to be allowed. 
However, these two worries are not necessary. 
As for the first one, we note that the fermion $\Psi$ in the operator (\ref{MMMD}) 
should be understood as the zero mode fermion. 
Though the operator (\ref{MMMD}) does not vanish even after imposing the on-shell condition, 
the remaining operator $M \varepsilon(y) \bar{\Psi} \Gamma^{MN} \Psi F_{MN}$ has no correspondence 
in the standard model (in the standard model $\bar{\Psi}_L \gamma^{MN} \Psi_R$ is not gauge invariant), 
and therefore vanishes automatically for the zero-mode fermion $\Psi$. 
As for the second one, note that the shift symmetry 
$A_y \to A_y + {\rm const}$ is operative as a remnant of higher dimensional gauge symmetry
even at the branes \cite{GIQ}. 
Therefore, the brane localized operator (\ref{braneAMM}) is forbidden by the shift symmetry. 
Furthermore, the UV finiteness is independent of how we compactify the extra space, 
because the information about the compactification is an infrared property of the theory. 
From these observations, we can expect the magnetic moment still to be finite 
even for the orbifold compactification and the presence of bulk mass term. 
The primary purpose of this paper is to clarify the cancellation mechanism of UV divergences 
in the calculation of the anomalous magnetic moment in the framework of the present orbifold model. 


This paper is organized as follows. 
In the next section, we introduce our model and discuss the mass eigenvalues and mode functions of 
fermions and gauge bosons. 
In section 3, we derive various interaction vertices, which are needed in the calculation of 
the anomalous magnetic moment and generally valid without any approximation. 
In section 4, we provide the general formulae for the contributions of $A_\mu$ ($D$-dimensional gauge field) 
and $A_y$ ($D$-dimensional scalar) exchange diagrams to the anomalous magnetic moment. 
The mechanism of cancellation of divergence is clarified in the case of small bulk mass in section 5. 
Our conclusions are given in section 6.  
The detailed derivation of some useful properties 
concerning the vertex functions are summarized in Appendices A and B.

\section{The Model}

We consider a $(D+1)$ dimensional $SU(3)$ gauge-Higgs unification model compactified on an orbifold $S^1/Z_2$ 
($S^1$: a circle of radius $R$) with a massive bulk fermion in the fundamental representation of $SU(3)$ gauge group. 
The Lagrangian is given by
\bea
{\cal L} &=& -\frac{1}{2} {\rm Tr} (F_{MN} F^{MN}) + \bar{\Psi} (iD\!\!\!\!/ - M \varepsilon(y)) \Psi
\label{Lagrangian1}
\eea
where the indices $M,N = 0,1,2,3 \cdots, D$, 
the $(D+1)$ dimensional gamma matrices are $\Gamma^M  = (\gamma^\mu, i \gamma^{D+1})~(\mu = 0,1,2,3, \cdots, D-1)$, 
\bea
&&F_{MN} = \partial_M A_N - \partial_N A_M - i g [A_M, A_N], \\
&&D\!\!\!\!/ = \Gamma^M (\partial_M -i g A_M), \\
&&\Psi = (\psi_1, \psi_2, \psi_3)^T. 
\eea
$g$ denotes a gauge coupling constant in $(D+1)$ dimensional gauge theory. 
$M$ is a bulk mass of the fermion. 
Here we note that the bulk fermion mass must have an odd $Z_2$ parity 
to be consistent with an orbifold projection: we thus introduce here the bulk mass 
proportional to the sign function $\varepsilon(y)$ of compactified extra space coordinate $y$. 

The periodic boundary condition is imposed along $S^1$ 
and $Z_2$ parity assignments are taken as
\bea
A_\mu = 
\left(
\begin{array}{ccc}
(+,+) & (+,+) & (-,-) \\
(+,+) & (+,+) & (-,-) \\
(-,-) & (-,-) & (+,+) \\
\end{array}
\right), \quad 
A_y = 
\left(
\begin{array}{ccc}
(-,-) & (-,-) & (+,+) \\
(-,-) & (-,-) & (+,+) \\
(+,+) & (+,+) & (-,-) \\
\end{array}
\right), 
\eea
\bea
\Psi = 
\left(
\begin{array}{c}
\psi_{1L}(+,+) + \psi_{1R}(-,-) \\
\psi_{2L}(+,+) + \psi_{2R}(-,-) \\
\psi_{3L}(-,-) + \psi_{3R}(+,+) \\
\end{array}
\right)
\eea
where $\mu = 0,1,2,3, \cdots, D-1$  
and $(+,+)$ means that $Z_2$ parities are even at the fixed points $y=0$ and $y=\pi R$, for instance.
$L, R$ on fermion denotes the chiral projection operator (for even $D$) defined as 
$L = \frac{1+\gamma^{D+1}}{2}, R = \frac{1-\gamma^{D+1}}{2}$. 
As can be seen from the KK mode expansion consistent with the boundary conditions 
only the fields with $(+,+)$ parities have massless modes as
\bea
A_\mu^{(0)} = \frac{1}{2}
\left(
\begin{array}{ccc}
W_\mu^3 +\frac{B_\mu}{\sqrt{3}} & \sqrt{2} W_\mu^+ & 0 \\
\sqrt{2} W_\mu^- & -W_\mu^3 + \frac{B_\mu}{\sqrt{3}} & 0 \\
0 & 0 & -\frac{2}{\sqrt{3}}B_\mu \\
\end{array}
\right), \quad 
A_y^{(0)} = \frac{1}{\sqrt{2}}
\left(
\begin{array}{ccc}
0 & 0 & \phi^+ \\
0 & 0 & \frac{h -i \phi^0}{\sqrt{2}} \\
\phi^- & \frac{h +i \phi^0}{\sqrt{2}} & 0 \\
\end{array}
\right), 
\label{0modegauge}
\eea
where $W_\mu^3, W_\mu^\pm$ and $B_\mu$ are the $SU(2)_L$ and $U(1)_Y$ gauge fields, respectively. 
From this expression, we see that the gauge symmetry $SU(3)$ is broken to $SU(2)_L \times U(1)_Y$ 
by the boundary conditions. 
Furthermore, the SM Higgs doublet is just embedded into the off-diagonal elements of  $A_y^{(0)}$. 
As for the fermion, we obtain massless modes 
\bea
\Psi^{(0)} = 
\left(
\begin{array}{c}
\psi_{1L}^{(0)} \\
\psi_{2L}^{(0)} \\
\psi_{3R}^{(0)} \\
\end{array}
\right) 
= 
\left(
\begin{array}{c}
u_L \\
d_L \\
d_R \\
\end{array}
\right)
\eea
which shows that chiral fermions are realized by orbifold projection. 


Some comments on this model are in order. 
First, the predicted Weinberg angle of this model is not realistic, $\sin^2 \theta_W = 3/4$. 
As was also discussed in \cite{GW}, 
the present $SU(3)$ model in five dimension ($D=4$) is inconsistent with 
the experimental requirements $\sin^2 \theta_W \simeq 1/4$ 
and $\rho = m_W^2/(m_Z^2\cos^2 \theta_W) =1$. 
Possible way to cure the problem is to introduce an extra U(1) or 
the brane localized gauge kinetic term \cite{SSS}. 
Second, the up quark remains massless and we have no up-type Yukawa coupling. 
A possible way out of this situation is to introduce second-rank symmetric tensors of $SU(3)$ 
(${\bf 6}$ dimensional representation) \cite{CCP}. 
Third point is that the fermion in the fundamental representation of $SU(3)$ has no lepton. 
In order to incorporate the leptons, a third-rank symmetric tensor 
(${\bf 10}$ dimensional representation) must be introduced. 
When such higher dimensional representations are added to the theory, there appear some massless exotic fermions and they should be removed from low energy sector 
by adding the brane localized fields to form brane localized mass terms with the exotic states. 

Since our primary purpose in this paper is to clarify 
the cancellation mechanism of UV divergence for the magnetic moment, 
in this paper we calculate the $g-2$ of down quark belonging to the triplet, as an example of the anomalous magnetic moment of fermion. This makes our calculation greatly simplified since the fundamental representation has no massless exotic fermions and we does not need introducing additional brane localized fermions and mass terms. We hope that our results for UV finiteness remains  unchanged even for the case of well-discussed muon $g-2$, since the operator analysis given in the introduction is independent of the representation of fermion. We also hope that our calculation of $g-2$ of $d$ quark will be relevant for, $e.g.,$ the electric dipole moment of neutron. 

Throughout this paper, what we mean by ``realistic" is in it's restricted sence, i.e. we mean that the gauge-Higgs model we consider contains the gauge group of the Standard Model and reproduces correct order of small Yukawa coupling relevant for the calculation of the magnetic moment.

\subsection{The mass eigenvalues and mode functions of fermions}

In order to derive $D$-dimensional effective Lagrangian and Feynman rules 
necessary for the calculations of the magnetic moment, 
first we have to obtain the $D$-dimensional mass eigenvalues and corresponding mode functions of  fermions. 

We first focus on the down quark sector $\psi_d \equiv (\psi_2, \psi_3)^t$ of the triplet fermion 
$(\psi_1, \psi_2, \psi_3)^t$. 
The $D$-dimensional mass term reads as 
\bea
{\cal L}_{{\rm mass}} &=& \bar{\psi}_d [\Gamma^y (i \partial_y + g \langle A_y \rangle) -M \varepsilon(y) ] \psi_d, 
\label{massfermion1} \\
g \langle A_y \rangle &\equiv& g_D v 
\left(
\begin{array}{cc}
0 & 1 \\
1 & 0 \\
\end{array}
\right) \quad 
\left( g_D \equiv \frac{g}{\sqrt{2\pi R}} \right).
\label{massfermion2}
\eea
We diagonalize the matrix $g \langle A_y \rangle$ by an orthogonal transformation, 
\bea
\tilde{\psi}_d = 
\left(
\begin{array}{c}
\tilde{\psi}_2 \\
\tilde{\psi}_3 \\
\end{array}
\right) 
\equiv  
\left(
\begin{array}{cc}
\frac{1}{\sqrt{2}} & \frac{1}{\sqrt{2}} \\
-\frac{1}{\sqrt{2}} & \frac{1}{\sqrt{2}} \\
\end{array}
\right) \psi_d 
= \frac{1}{\sqrt{2}} \left(
\begin{array}{c}
\psi_2+\psi_3 \\
-\psi_2+\psi_3 \\
\end{array}
\right). 
\eea
In terms of $\tilde{\psi}_{2,3}$, (\ref{massfermion1}) can be rewritten as
\bea
\bar{\tilde{\psi}_2} [\Gamma^y (i \partial_y + g_D v) - M \varepsilon(y) ] \tilde{\psi}_2 
+ \bar{\tilde{\psi}_3} [\Gamma^y (i \partial_y - g_D v) - M \varepsilon(y) ] \tilde{\psi}_3. 
\label{massfermion3}
\eea
Now, we try to find mass eigenvalues and mode functions of $\tilde{\psi}_2$. 
Expanding the $\tilde{\psi}_2(x,y)$ as 
$\tilde{\psi}_2(x,y) = \sum_n \frac{1}{\sqrt{2}} [ d_{L}^{(n)}(x) f_{d_L}^{(n)}(y) + d_{R}^{(n)}(x) f_{d_R}^{(n)}(y) ]$, 
we found the equations of motion 
\bea
&&[ (i \partial_y + g_D v)^2 + M^2 -2M (\delta(y) - \delta (y-\pi R) ) ] f_{d_L}^{(n)}(y) = m_n^2 f_L^{(n)}(y), 
\label{modeeqL}\\
&&[ (i \partial_y + g_D v)^2 + M^2 +2M (\delta(y) - \delta (y-\pi R) ) ] f_{d_R}^{(n)}(y) = m_n^2 f_R^{(n)}(y)
\label{modeeqR}
\eea 
where $m_n$ is the mass eigenvalue of $n$-th KK mode with $n$ being an arbitrary integer. 

Let us solve (\ref{modeeqL}). 
By use of the boundary condition that $f_{d_L}^{(n)}$ is continuous 
and $\partial_y f_{d_L}^{(n)}$ has a discontinuity $-2M f_{d_L}^{(n)}$ 
at the fixed point $y=0$, 
the solution is known to take a form 
\bea
f_{d_L}^{(n)}(y) &=& e^{ig_D v y} \left[ 
C_1^{(n)} \left\{ \cos(\sqrt{m_n^2-M^2}y) 
- \frac{M}{\sqrt{m_n^2-M^2}} \varepsilon(y) \sin(\sqrt{m_n^2-M^2}y) \right\} \right. \nonumber \\
&& \left. + C_2^{(n)} \sin(\sqrt{m_n^2-M^2}y) \right] 
\label{generalsolL}
\eea
where $C_{1,2}^{(n)}$ are integration constants. 

Then the similar boundary conditions 
at the fixed point $y=\pi R$ can be written as
\bea
&&0 = (w-1) \left(\cos \varphi_n -\frac{M}{\sqrt{m_n^2-M^2}} \sin \varphi_n \right) C_1^{(n)} 
+ (w+1) \sin \varphi_n C_2^{(n)}, 
\label{bc1} \\
&&0 = -(w+1) \frac{m_n^2}{m_n^2-M^2} \sin \varphi_n C_1^{(n)} 
+ (w-1) \left(\cos \varphi_n + \frac{M}{\sqrt{m_n^2-M^2}} \sin \varphi_n \right) C_2^{(n)} \nonumber 
\label{bc2} \\
\eea
where $w$ is a Wilson loop $w \equiv e^{2\pi i g_D Rv}$ and $\varphi_n \equiv \sqrt{m_n^2-M^2} \pi R$. 
These boundary conditions determine the mass eigenvalues $m_n$ through the condition
\bea
\frac{m_n^2}{m_n^2-M^2} \sin^2(\sqrt{m_n^2-M^2}\pi R) = \sin^2(g_D v \pi R), 
\label{masscond}
\eea
which cannot be solved analytically for $m_n$, in general.  
As a check, 
if we consider the case of $v=0$, we obtain
\bea
m_0 = 0, \quad m_n^2 = \left( \frac{n}{R} \right)^2 + M^2 
\eea
which is a well known result. 
We take a sign convention such that  
\bea
m_n = \frac{n}{R} + g_D v 
\eea
in the case of $M=0$. 
For $m_n$ satisfying (\ref{masscond}), the ratio of $C_1^{(n)}$ and $C_2^{(n)}$ is known to be fixed as
\bea
C_1^{(n)} / C_2^{(n)} = \sqrt{\cos(\varphi_n-\alpha_n)} / -i \frac{m_n}{\sqrt{m_n^2 - M^2}} \sqrt{\cos(\varphi_n + \alpha_n)}
\eea
where $\tan \alpha_n \equiv \frac{M}{\sqrt{m_n^2-M^2}}$. 
Thus, we have obtained the mode functions
\bea
&&f_{d_L}^{(n)}(y) = F_{M, M_W}^{(n)}(y), \quad f_{d_R}^{(n)}(y) = F_{-M, M_W}^{(n)}(y), \\
&&F_{M,M_W}^{(n)}(y) = e^{ig_D vy}C^{(n)} 
\left[
\sqrt{\cos(\varphi_n -\alpha_n)} \cos(\sqrt{m_n^2-M^2}|y| + \alpha_n) \right. \nonumber \\
&& \left. \hspace*{4cm} - \varepsilon(n) i \sqrt{\cos(\varphi_n + \alpha_n)} \sin(\sqrt{m_n^2-M^2}y) 
\right], \label{modeL}
\label{modeR}
\eea
where the normalization constant $C^{(n)}$ is given by
\bea
C^{(n)} &=& \left[
2\pi R \cos \varphi_n \cos \alpha_n  
- \frac{2}{\sqrt{m_n^2-M^2}} \sin \varphi_n  \cos \alpha_n \sin^2 \alpha_n 
\right]^{-1/2}. 
\label{normalization}
\eea
The ``sign function" $\varepsilon(n)$ is defined as $1$ for $n \ge 0$ and $-1$ for $n < 0$. 

As a matter of fact when the mode functions are substituted in (\ref{massfermion3}) 
we get a mass $-i m_n$ for the Dirac fermion $d^{(n)} = d_L^{(n)} + d_R^{(n)}$. 
Thus we perform a chiral transformation
\bea
\hat{\psi}_d \equiv 
\left(
\begin{array}{c}
\hat{\psi}_2 \\
\hat{\psi}_3 \\
\end{array}
\right)
= e^{-i\frac{\pi}{4}\gamma^{D+1}} \tilde{\psi}_2,
\eea
so that $d^{(n)}$ has a mass $m_n$, with $\hat{\psi}_2$ still being mode-expanded as 
\bea
\hat{\psi}_2(x,y) = \sum_{n=-\infty}^\infty \frac{1}{\sqrt{2}} 
\left(
f_{d_L}^{(n)} d_L^{(n)} + f_{d_R}^{(n)} d_R^{(n)}
\right). 
\label{modeexpsi2}
\eea

Because of the orbifolding, 
$\hat{\psi}_3$ is not independent of $\hat{\psi}_2$. 
In fact, the $Z_2$ parity assignment $\psi_2(x,-y) = \gamma^{D+1} \psi_2(x,y)$ and 
$\psi_3(x,-y) = -\gamma^{D+1} \psi_3(x,y)$ tells us
\bea
\hat{\psi}_3 (x,y) = -\gamma^{D+1} \hat{\psi}_2(x,-y). 
\eea
Since $\psi_1$ does not get a mass due to the VEV $v$ and the $Z_2$ parity assignment is 
$\psi_1(x,-y) = \gamma^{D+1} \psi_1(x,y)$, $\hat{\psi}_1 \equiv e^{-i\frac{\pi}{4}\gamma^{D+1}} \psi_1$ 
is mode-expanded as
\bea
\hat{\psi}_1(x,y) &=& \sum_{n=1}^\infty 
\left\{
\frac{1}{\sqrt{2}} (f_{u_L}^{(n)}(y) + f_{u_L}^{(n)}(-y)) u_L^{(n)}(x) 
+ \frac{1}{\sqrt{2}} (f_{u_R}^{(n)}(y) - f_{u_R}^{(n)}(-y)) u_R^{(n)}(x)
\right\} \nonumber \\
&& + f_{u_L}^{(0)}(y) u_L^{(0)}(x),  
\eea
where the mode functions take relatively simple forms
\bea
f_{u_L}^{(n)}(y) &=& F_{M, 0}^{(n)}(y) = \frac{1}{\sqrt{2\pi R}} 
\left[
\cos \left(\frac{n}{R} |y| + \alpha_n^{(0)} \right) -i \sin \left(\frac{n}{R} y \right)
\right], \\
f_{u_L}^{(0)}(y) &=& F_{M, 0}^{(0)}(y) = \sqrt{\frac{M}{1-e^{-2\pi MR}}} e^{-M|y|}, \\ 
f_{u_R}^{(n)}(y) &=& F_{-M, 0}^{(n)}(y) = \frac{1}{\sqrt{2\pi R}} 
\left[
\cos \left(\frac{n}{R} |y| - \alpha_n^{(0)} \right) -i \sin \left(\frac{n}{R} y \right)
\right], 
\eea
where $\alpha_n^{(0)}$ is defined by
\bea
\tan \alpha_n^{(0)} \equiv \frac{M}{n/R}~(n \ge 1). 
\eea
Dirac fermion $u^{(n)}(x) \equiv u_L^{(n)}(x) + u_R^{(n)}(x)$ has a mass
\bea
\tilde{m}_n = \sqrt{\left(\frac{n}{R} \right)^2 + M^2}~(n \ge 1), 
\eea
while $u_L^{(0)}(x)$ remains as a massless state: $\tilde{m}_0 = 0$. 

We thus have obtained the mass eigenstates of the fermion, 
\bea
\hat{\Psi} &\equiv& 
\left(
\begin{array}{c}
\hat{\psi}_1 \\
\hat{\psi}_2 \\
\hat{\psi}_3 \\
\end{array}
\right) 
= e^{-i \frac{\pi}{4}\gamma_y} {\cal O} \Psi \nonumber \\
&=& 
\left(
\begin{array}{c}
\sum_{n=1}^\infty \frac{1}{\sqrt{2}} 
\left\{
(f_{u_L}^{(n)}(y) + f_{u_L}^{(n)}(-y)) u_L^{(n)}(x) 
+ (f_{u_R}^{(n)}(y) - f_{u_R}^{(n)}(-y)) u_R^{(n)}(x)
\right\}
+ f_{u_L}^{(0)}(y) u_L^{(0)}(x) \\
\sum_{n=-\infty}^\infty \frac{1}{\sqrt{2}} (f_{d_L}^{(n)}(y) d_L^{(n)}(x) + f_{d_R}^{(n)}(y) d_R^{(n)}(x)) \\
\sum_{n=-\infty}^\infty \frac{1}{\sqrt{2}} (-f_{d_L}^{(n)}(-y) d_L^{(n)}(x) + f_{d_R}^{(n)}(-y) d_R^{(n)}(x)) \\
\end{array}
\right), \nonumber \\
\eea
where
\bea
{\cal O} &\equiv& 
\left(
\begin{array}{ccc}
1 & 0 & 0 \\
0 & \frac{1}{\sqrt{2}} & \frac{1}{\sqrt{2}} \\
0 & -\frac{1}{\sqrt{2}} & \frac{1}{\sqrt{2}} \\
\end{array}
\right). 
\label{rotation}
\eea
Let us comment on the quantum mechanical supersymmetry (QMS) hidden in the mode functions of fermions. 
It has been demonstrated that the mode functions of $A_\mu$ and $A_y$ form a supermultiplet 
for each non-zero KK modes, which reflects the Higgs-like mechanism to form massive gauge bosons 
$A_\mu^{(n)}(n \ne 0)$ \cite{LNSS}. 
A similar thing is expected to take place in the fermion sector; 
$f_L^{(n)}$ and $f_R^{(n)}$ are expected to form a supermultiplet, 
as they are ``partner" to form a massive Dirac fermion. 
In fact, they are known to be related by a supercharge $Q$: 
\bea
&&Q
\left(
\begin{array}{c}
F_{M, M_W}^{(n)} \\
F_{-M, M_W}^{(n)} \\
\end{array}
\right)
= m_n 
\left(
\begin{array}{c}
F_{M, M_W}^{(n)} \\
F_{-M, M_W}^{(n)} \\
\end{array}
\right), 
\label{2.35}\\
&&Q = 
\left(
\begin{array}{cc}
0 & i \partial_y + M_W -i M \varepsilon(y) \\
i \partial_y + M_W +i M \varepsilon(y) & 0 \\
\end{array}
\right)
\label{2.36}
\eea
with $Q^2$ giving the differential operators in (\ref{modeeqL}) and (\ref{modeeqR}), 
namely the Hamiltonian in QMS. 

\subsection{The mass eigenvalues and mode functions of gauge bosons}
Next, we turn to the mass eigenvalues and mode functions of $D$-dimensional gauge bosons and Higgs scalars, 
$i.e.~A_\mu$ and $A_y$. 
First we explicitly write the gauge bosons and Higgs scalar fields as follows, 
\bea
A_\mu(x, y) &=& 
\left(
\begin{array}{ccc}
\frac{\gamma_\mu}{\sqrt{3}} & \frac{W^+_\mu}{\sqrt{2}} & \frac{\phi^+_\mu}{\sqrt{2}} \\
\frac{W^-_\mu}{\sqrt{2}} & -\frac{\gamma_\mu}{2\sqrt{3}} -\frac{Z_\mu}{2} & \frac{h_\mu-i\phi^0_\mu}{2} \\
\frac{\phi^-_\mu}{\sqrt{2}} & \frac{h_\mu +i \phi^0_\mu}{2} & -\frac{\gamma_\mu}{2\sqrt{3}} + \frac{Z_\mu}{2} \\
\end{array}
\right), 
\label{gaugematrix} \\
A_y(x, y) &=& 
\left(
\begin{array}{ccc}
\frac{\gamma_y}{\sqrt{3}} & \frac{W^+_y}{\sqrt{2}} & \frac{\phi^+}{\sqrt{2}} \\
\frac{W^-_y}{\sqrt{2}} & -\frac{\gamma_y}{2\sqrt{3}} -\frac{Z_y}{2} & \frac{h-i\phi^0}{2} \\
\frac{\phi^-}{\sqrt{2}} & \frac{h +i \phi^0}{2} & -\frac{\gamma_y}{2\sqrt{3}} + \frac{Z_y}{2} \\
\end{array}
\right). 
\label{Aymatrix}
\eea
Each field has a mode expansion depending on its $Z_2$ parity, 
\bea
A_{\mu,y}(x,y) &=& \frac{1}{\sqrt{2\pi R}} A_{\mu,y}^{(0)}(x) 
+ \frac{1}{\sqrt{\pi R}} \sum_{n=1}^\infty A_{\mu,y}^{(n)}(x) \cos \left(\frac{n}{R}y \right)~({\rm even}), \\
A_{\mu,y}(x,y) &=&  
\frac{1}{\sqrt{\pi R}} \sum_{n=1}^\infty A_{\mu,y}^{(n)}(x) \sin \left(\frac{n}{R}y \right)~({\rm odd})
\eea
Putting these mode functions into the term ${\rm Tr}(F_{\mu y})^2$ and integrating over $y$ coordinate lead to 
the necessary quadratic terms concerning nonzero KK modes;
\bea
{\cal L}_{{\rm quadratic}} &=& \sum_{n=1}^\infty \left[ 
\frac{1}{2} \left(\partial_\mu \gamma_y^{(n)} + \frac{n}{R} \gamma_\mu^{(n)} \right)^2 
+ \frac{1}{2} \left(\partial_\mu h^{(n)} - \frac{n}{R} h_\mu^{(n)} \right)^2 
+ \frac{1}{2}[(\partial_\mu Z_y^{(n)})^2 + (\partial_\mu \phi^{0(n)})^2] \right. \nonumber \\
&& \left. + \frac{1}{2}\left[ \left( \frac{n}{R} \right)^2 + (2g_D v)^2 \right] [ (Z_\mu^{(n)})^2 + (\phi_\mu^{0(n)})^2 ]
+ 4\frac{n}{R} g_D v Z_\mu^{(n)} \phi^{\mu0(n)} \right. \nonumber \\
&& \left. + (\partial^\mu Z_y^{(n)}) \left(\frac{n}{R} Z_\mu^{(n)} +2g_D v \phi_\mu^{0(n)} \right) 
- (\partial^\mu \phi^{0(n)}) \left(\frac{n}{R} \phi_\mu^{0(n)} +2g_D v Z_\mu^{(n)} \right) \right. \nonumber \\
&& \left. +|\partial_\mu W_y^{+(n)}|^2 + |\partial_\mu \phi^{+(n)}|^2 
+ \left[\left( \frac{n}{R} \right)^2 + (g_D v)^2 \right](|W_\mu^{+(n)}|^2 + |\phi_\mu^{+(n)}|^2) \right. \nonumber \\
&& \left. + \left\{
i\frac{n}{R} 2g_D v W_\mu^{+(n)} \phi^{-(n)\mu} 
+(\partial^\mu W_y^{+(n)}) \left(\frac{n}{R} W_\mu^{-(n)} +ig_D v \phi_\mu^{-(n)} \right) \right. \right. \nonumber \\
&& \left. \left. 
+(\partial^\mu \phi^{+(n)}) \left(-\frac{n}{R} \phi_\mu^{-(n)} +ig_D v W_\mu^{-(n)} \right) + {\rm h.c.}
\right\} \right]
\eea
The mixings between $Z$ and $\phi^0$, $W^\pm$ and $\phi^\pm$ due to the VEV $v$ 
necessitates the following diagonalization,
\bea
&&\left(
\begin{array}{c}
Z_{1\mu}^{(n)} \\
Z_{2\mu}^{(n)} \\
\end{array}
\right)
= \left(
\begin{array}{cc}
\frac{1}{\sqrt{2}} & \frac{1}{\sqrt{2}} \\
\frac{1}{\sqrt{2}} & -\frac{1}{\sqrt{2}} \\
\end{array}
\right)
\left(
\begin{array}{c}
Z_{\mu}^{(n)} \\
\phi_{\mu}^{0(n)} \\
\end{array}
\right), \quad 
\left(
\begin{array}{c}
Z_{1y}^{(n)} \\
Z_{2y}^{(n)} \\
\end{array}
\right)
= \left(
\begin{array}{cc}
\frac{1}{\sqrt{2}} & -\frac{1}{\sqrt{2}} \\
\frac{1}{\sqrt{2}} & \frac{1}{\sqrt{2}} \\
\end{array}
\right)
\left(
\begin{array}{c}
Z_y^{(n)} \\
\phi^{0(n)} \\
\end{array}
\right), 
\label{rotationZ}\\
&&\left(
\begin{array}{c}
W_{1\mu}^{+(n)} \\
W_{2\mu}^{+(n)} \\
\end{array}
\right)
= \left(
\begin{array}{cc}
\frac{1}{\sqrt{2}} & -\frac{1}{\sqrt{2}} \\
\frac{1}{\sqrt{2}} & \frac{1}{\sqrt{2}} \\
\end{array}
\right)
\left(
\begin{array}{c}
W_{\mu}^{+(n)} \\
i \phi_{\mu}^{+(n)} \\
\end{array}
\right), \quad 
\left(
\begin{array}{c}
W_{1y}^{+(n)} \\
W_{2y}^{+(n)} \\
\end{array}
\right)
= \left(
\begin{array}{cc}
\frac{1}{\sqrt{2}} & \frac{1}{\sqrt{2}} \\
\frac{1}{\sqrt{2}} & -\frac{1}{\sqrt{2}} \\
\end{array}
\right)
\left(
\begin{array}{c}
W_y^{+(n)} \\
i \phi^{+(n)} \\
\end{array}
\right). \nonumber 
\label{rotationW}\\
\eea
Thus, the quadratic terms including the zero mode sector read as
\bea
{\cal L}_{{\rm quadratic}} &=& \sum_{n=1}^\infty \left[ 
\frac{1}{2} \left(\partial_\mu \gamma_y^{(n)} + \frac{n}{R} \gamma_\mu^{(n)} \right)^2 
+ \frac{1}{2} \left(\partial_\mu h^{(n)} - \frac{n}{R} h_\mu^{(n)} \right)^2 \right. \nonumber \\
&& \left. 
+ \frac{1}{2} \left[ \partial_\mu Z_{1y}^{(n)} + \left(\frac{n}{R} + M_Z \right) Z_{1\mu}^{(n)} \right]^2 
+ \frac{1}{2} \left[ \partial_\mu Z_{2y}^{(n)} + \left(\frac{n}{R} - M_Z \right) Z_{2\mu}^{(n)} \right]^2 \right. \nonumber \\
&& \left. + \left|\partial_\mu W_{1y}^{+(n)} + \left(\frac{n}{R} + M_W \right) W_{1\mu}^{+(n)} \right|^2 
+ \left|\partial_\mu W_{2y}^{+(n)} + \left(\frac{n}{R} - M_W \right) W_{2\mu}^{+(n)} \right|^2 \right] \nonumber \\
&& 
+ \frac{1}{2}(\partial_\mu h^{(0)})^2 + \frac{1}{2}(\partial_\mu \phi^{0(0)} + M_Z Z_\mu^{(0)})^2 
+|\partial_\mu (i\phi^{+(0)}) +M_W W_\mu^{+(0)}|^2
\label{gaugemass}
\eea
where $M_Z = 2g_D v =2M_W$ (recall that $\sin \theta_W = \frac{\sqrt{3}}{2}$). 

We note that the Higgs-like mechanism works between the partner of the pairs $(Z_{1,2\mu}^{(n)}, Z_{1,2y}^{(n)})$ 
and $(W_{1,2\mu}^{+(n)}, W_{1,2y}^{+(n)})$. 
As we will see later, this structure is crucial for the cancellation of UV divergences 
in the calculation of the anomalous moment. 
The mass spectrum for degenerate pairs of the gauge bosons and would-be N-G bosons 
(in the 't Hooft-Feynman gauge) are summarized as
\bea
(Z_{1\mu}^{(n)}, Z_{1y}^{(n)}) &:& \frac{n}{R} + M_Z, \\
(Z_{2\mu}^{(n)}, Z_{2y}^{(n)}) &:& \frac{n}{R} - M_Z, \\
(W_{1\mu}^{+(n)}, W_{1y}^{+(n)}) &:& \frac{n}{R} + M_W, \\
(W_{2\mu}^{+(n)}, W_{2y}^{+(n)}) &:& \frac{n}{R} - M_W.  
\eea

We thus find that just as in the sector of fermions, the mass eigenstates are obtained 
by the orthogonal transformation due to ${\cal O}$ in (\ref{rotation}): 
\bea
\tilde{A}_\mu &\equiv& {\cal O} A_\mu {\cal O}^t 
= 
\left(
\begin{array}{ccc}
\frac{\gamma_\mu}{\sqrt{3}} & \frac{W_\mu^+ + \phi_\mu^+}{2} & \frac{-W_\mu^+ + \phi_\mu^+}{2} \\
\frac{W_\mu^- + \phi_\mu^-}{2} & -\frac{\gamma_\mu}{2 \sqrt{3}} + \frac{h_\mu}{2} & \frac{Z_\mu -i \phi_\mu^0}{2} \\
\frac{-W_\mu^- + \phi_\mu^-}{2} & \frac{Z_\mu +i \phi_\mu^0}{2} & -\frac{\gamma_\mu}{2 \sqrt{3}} - \frac{h_\mu}{2} \\
\end{array}
\right) \nonumber \\
&=& 
\left(
\begin{array}{cc}
\frac{1}{\sqrt{3}} \left(\gamma^{(n)}_\mu C_n + \frac{\gamma_\mu^{(0)}}{\sqrt{2\pi R}} \right) & 
\frac{1}{2} \hat{W}_\mu^{+(n)} \frac{e^{i\frac{n}{R}y}}{\sqrt{2\pi R}} \\
\frac{1}{2} \hat{W}_\mu^{-(n)} \frac{e^{-i\frac{n}{R}y}}{\sqrt{2\pi R}} & 
-\frac{1}{2 \sqrt{3}} \left(\gamma^{(n)}_\mu C_n + \frac{\gamma_\mu^{(0)}}{\sqrt{2\pi R}} \right) + \frac{1}{2}h_\mu^{(n)}S_n \\
-\frac{1}{2} \hat{W}_\mu^{-(n)} \frac{e^{i\frac{n}{R}y}}{\sqrt{2\pi R}} & \frac{1}{2}\hat{Z}_\mu^{(n)} \frac{e^{i\frac{n}{R}y}}{\sqrt{2\pi R}} \\
\end{array}
\right. \nonumber \\
&& \left.
\begin{array}{c}
-\frac{1}{2} \hat{W}_\mu^{+(n)} \frac{e^{-i\frac{n}{R}y}}{\sqrt{2\pi R}} \\
\frac{1}{2}\hat{Z}_\mu^{(n)} \frac{e^{-i\frac{n}{R}y}}{\sqrt{2\pi R}} \\
-\frac{1}{2 \sqrt{3}} \left(\gamma^{(n)}_\mu C_n + \frac{\gamma_\mu^{(0)}}{\sqrt{2\pi R}} \right) - \frac{1}{2}h_\mu^{(n)} S_n \\
\end{array}
\right), 
\label{gaugefield1} \\
\tilde{A}_y &\equiv& {\cal O} A_y {\cal O}^t 
= 
\left(
\begin{array}{ccc}
\frac{\gamma_y}{\sqrt{3}} & \frac{W_y^+ + \phi^+}{2} & \frac{-W_y^+ + \phi^+}{2} \\
\frac{W_y^- + \phi^-}{2} & -\frac{\gamma_y}{2 \sqrt{3}} + \frac{h}{2} & \frac{Z_y -i \phi^0}{2} \\
\frac{-W_y^- + \phi^-}{2} & \frac{Z_y +i \phi^0}{2} & -\frac{\gamma_y}{2 \sqrt{3}} - \frac{h}{2} \\
\end{array}
\right) \nonumber \\
&=& 
\left(
\begin{array}{cc}
\frac{1}{\sqrt{3}} \gamma^{(n)}_y S_n & 
\frac{1}{2} \hat{\phi}^{+(n)} \frac{e^{i\frac{n}{R}y}}{\sqrt{2\pi R}} \\
\frac{1}{2} \hat{\phi}^{-(n)} \frac{e^{-i\frac{n}{R}y}}{\sqrt{2\pi R}} & 
-\frac{1}{2 \sqrt{3}} \gamma^{(n)}_y S_n + \frac{1}{2} \left( h^{(n)} C_n + \frac{h^{(0)}}{\sqrt{2\pi R}} \right) \\
\frac{1}{2} \hat{\phi}^{-(n)} \frac{e^{i\frac{n}{R}y}}{\sqrt{2\pi R}} & \frac{i}{2} \hat{\phi}^{0(n)} \frac{e^{i\frac{n}{R}y}}{\sqrt{2\pi R}} \\
\end{array}
\right. \nonumber \\
&& \left.
\begin{array}{c}
\frac{1}{2} \hat{\phi}^{+(n)} \frac{e^{-i\frac{n}{R}y}}{\sqrt{2\pi R}} \\
-\frac{i}{2}\hat{\phi}^{0(n)} \frac{e^{-i\frac{n}{R}y}}{\sqrt{2\pi R}} \\
-\frac{1}{2 \sqrt{3}} \gamma^{(n)}_y S_n -\frac{1}{2} \left( h^{(n)} C_n + \frac{h^{(0)}}{\sqrt{2\pi R}} \right) \\
\end{array}
\right), 
\eea
where $C_n \equiv \frac{\cos \left(\frac{n}{R}y \right)}{\sqrt{\pi R}}, S_n \equiv \frac{\sin \left(\frac{n}{R}y \right)}{\sqrt{\pi R}}$, 
and the mode sum is for $n \ge 1$ in the case of $C_n$ and $S_n$, while the sum is for all integer 
in the case of $e^{\pm i \frac{n}{R} y}/\sqrt{2\pi R}$. 
We have used the following notation, combining $Z_{1\mu,y}^{(n)}, W_{1\mu,y}^{+(n)}$ with $Z_{2\mu,y}^{(n)}, W_{2\mu,y}^{+(n)}$, 
respectively: 
\bea
&& \hat{Z}_\mu^{(n)} = Z_{1\mu}^{(n)}, \hat{Z}_\mu^{(-n)} = Z_{2\mu}^{(n)}~(n \ge 1), \hat{Z}_\mu^{(0)} = Z_\mu^{(0)} \\
&& \hat{\phi}^{0(n)} = -Z_{1y}^{(n)}, \hat{\phi}^{0(-n)} = Z_{2y}^{(n)}~(n \ge 1), \hat{\phi}^{0(0)} = \phi^{0(0)} \\
&& \hat{W}_\mu^{+(n)} = W_{1\mu}^{+(n)}, \hat{W}_\mu^{+(-n)} = W_{2\mu}^{+(n)}~(n \ge 1), \hat{W}_\mu^{+(0)} = W_\mu^{+(0)} \\
&& \hat{\phi}^{+(n)} = -i W_{1y}^{+(n)}, \hat{\phi}^{+(-n)} = i W_{2y}^{+(n)}~(n \ge 1), \hat{\phi}^{+(0)} = \phi^{+(0)},
\eea
so that the mass squared of $(\hat{Z}_\mu^{(n)}, \hat{\phi}^{0(n)})$ and $(\hat{W}_\mu^{+(n)}, \hat{\phi}^{+(n)})$ 
are $\left(\frac{n}{R} + M_Z \right)^2$ and $\left(\frac{n}{R} + M_W \right)^2$, respectively. 

\section{Interaction vertices}
Having obtained the mode functions of the fermion $\hat{\Psi}$ and the ``gauge-Higgs" sector $\tilde{A}_\mu, \tilde{A}_y$, 
we are ready to derive $D$-dimensional gauge and Yukawa interaction vertices of $\hat{\Psi}$ 
by the overlap integral of the relevant mode functions with respect to $y$, 
in the relevant part of the Lagrangian, 
\bea
g \bar{\hat{\Psi}} [\tilde{A}_\mu \gamma^\mu - \tilde{A}_y]\hat{\Psi}. 
\eea

Let us start with the interaction vertex of $\gamma_\mu$, with $d$ quark, 
which is obtained by an integral
\bea
&&-\frac{g}{2\sqrt{3}} \int_{-\pi R}^{\pi R} dy \gamma_\mu(x,y) 
\left[
\bar{\hat{\psi}}_2(x,y) \gamma^\mu \hat{\psi}_2(x,y) + \bar{\hat{\psi}}_3(x,y) \gamma^\mu \hat{\psi}_3(x,y) 
\right] \nonumber \\
&=& -\frac{g}{\sqrt{3}} \int_{-\pi R}^{\pi R}dy \gamma_\mu(x,y) \bar{\hat{\psi}}_2(x,y) \gamma^\mu \hat{\psi}_2(x,y) 
\label{Intvtx}
\eea
where $\gamma_\mu(x,y)$ denotes the photon field and 
the properties $\hat{\psi}_3(x,y) = -\gamma^{D+1} \hat{\psi}_2(x,-y)$ 
and $\gamma_\mu(x,-y) = \gamma_\mu(x,y)$ are used. 
Substituting (\ref{modeexpsi2}) and the $\gamma_\mu$ part of (\ref{gaugefield1}) in (\ref{Intvtx}), 
we get the $D$-dimensional vertex function necessary for the calculation of $g-2$ of the zero-mode $d^{(0)}$, 
\bea
-\frac{g}{2\sqrt{3}} \gamma_\mu^{(m)} \bar{d}^{(n)} \gamma^\mu V_{\gamma_\mu}^{nm} d^{(0)},
\label{3.3}
\eea
where the vertex function $V_{\gamma_\mu}^{nm}$ is defined as
\bea
V_{\gamma_\mu}^{nm} &=& I_c^{nm} (M, M_W,; M, M_W)L + I_c^{nm} (-M, M_W,; -M, M_W)R \nonumber \\
&=& (L + (-1)^{n+m}R) I_c^{nm}(M, M_W; M, M_W)~(m \ge 1), \label{3.4} \\
V_{\gamma_\mu}^{n0} &=& (L + (-1)^{n}R) I^{n0}(M, M_W; M, M_W)
\eea
in terms of functions
\bea
&&I_c^{nm}(M_1, M_2; M_3, M_4) \equiv \frac{1}{\sqrt{\pi R}} \int_{-\pi R}^{\pi R} dy 
\cos \left(\frac{m}{R}y \right) F_{M_1, M_2}^{(n)}(y)^* F_{M_3, M_4}^{(0)}(y)~(m \ge 1), 
\label{3.6}\\
&&I^{n0}(M_1, M_2; M_3, M_4) \equiv \frac{1}{\sqrt{2\pi R}} \int_{-\pi R}^{\pi R} dy 
F_{M_1, M_2}^{(n)}(y)^* F_{M_3, M_4}^{(0)}(y). 
\label{3.7}
\eea
Here we have used the fact that the vertices for the left-handed and right-handed $d$ quark are related 
with the factor $(-1)^{m+n}$, as is shown in the Appendix A. 
For the case of $M_1=M_3=M$ and $M_2=M_4=M_W$, the explicit forms of these functions are
\bea
&&I_c^{nm}(M, M_W; M, M_W) 
= \frac{1}{\sqrt{\pi R}} C^{(n)}(\varphi_n, \alpha_n)^* C^{(0)}(\varphi_0, \alpha_0) 
\left\{
\sqrt{\cos(\varphi_n-\alpha_n)} \sqrt{\cos(\varphi_0 -\alpha_0)} \right. \nonumber \\
&& \left. \left[
-\frac{\sqrt{m_n^2-M^2} + \sqrt{m_0^2-M^2}}{(\frac{m}{R})^2-(\sqrt{m_n^2-M^2} + \sqrt{m_0^2-M^2})^2} 
\left(
(-1)^m \sin(\varphi_n + \alpha_n + \varphi_0 + \alpha_0) - \sin(\alpha_n + \alpha_0)
\right) \right. \right. \nonumber \\
&& \left. \left.  
-\frac{\sqrt{m_n^2-M^2} - \sqrt{m_0^2-M^2}}{(\frac{m}{R})^2-(\sqrt{m_n^2-M^2} - \sqrt{m_0^2-M^2})^2} 
\left(
(-1)^m \sin(\varphi_n + \alpha_n - \varphi_0 - \alpha_0) - \sin(\alpha_n - \alpha_0)
\right)
\right] \right. \nonumber \\
&& \left. + \varepsilon(n) 
\sqrt{\cos(\varphi_n+\alpha_n)} \sqrt{\cos(\varphi_0 + \alpha_0)} \right. \nonumber \\
&& \left. \left[
\frac{\sqrt{m_n^2-M^2} + \sqrt{m_0^2-M^2}}{(\frac{m}{R})^2-(\sqrt{m_n^2-M^2} + \sqrt{m_0^2-M^2})^2} 
\left(
(-1)^m \sin(\varphi_n + \varphi_0) 
\right) \right. \right. \nonumber \\
&& \left.  \left. 
-\frac{\sqrt{m_n^2-M^2} - \sqrt{m_0^2-M^2}}{(\frac{m}{R})^2-(\sqrt{m_n^2-M^2} - \sqrt{m_0^2-M^2})^2} 
\left(
(-1)^m \sin(\varphi_n - \varphi_0) 
\right)
\right]
\right\}, \label{Icnm}\\
&& I^{n0}(M, M_W; M, M_W) = \delta_{n0} \frac{1}{\sqrt{2\pi R}}. 
\label{In0} 
\eea
Note that (\ref{In0}) is nothing but the orthonormality of the mode functions. 
We therefore find that the $\gamma_\mu^{(0)}$ coupling is of ordinary form: 
\bea
-\frac{g}{2\sqrt{3}} \frac{1}{\sqrt{2\pi R}} \gamma_\mu^{(0)} \bar{d}^{(0)} \gamma^\mu d^{(0)} 
= -\frac{e_D}{3} \gamma_\mu^{(0)} \bar{d}^{(0)} \gamma^\mu d^{(0)}, 
\eea
where $\frac{g}{\sqrt{2\pi R}} =g_D, g_D \sin \theta_W = \frac{\sqrt{3}}{2} g_D = e_D$ are used. 

The interaction vertex of $\gamma_y$ with $d$ quark is derived by a similar step 
as in the case of $\gamma_\mu$ vertex: 
\bea
&&\frac{g}{2\sqrt{3}} \int_{-\pi R}^{\pi R} dy \gamma_y(x,y) 
\left[
\bar{\hat{\psi}}_2(x,y) \hat{\psi}_2(x,y) + \bar{\hat{\psi}}_3(x,y) \hat{\psi}_3(x,y)
\right] \nonumber \\
&=& \frac{g}{\sqrt{3}} \int_{-\pi R}^{\pi R} dy \gamma_y(x,y) \bar{\hat{\psi}}_2(x,y) \hat{\psi}_2(x,y) 
\to \frac{g}{2\sqrt{3}} \gamma^{(m)}_y \bar{d}^{(n)} V_{\gamma_y}^{nm} d^{(0)}~(m \ge 1), 
\eea
where the vertex function $V_{\gamma_y}^{nm}$ is defined as 
\bea
V_{\gamma_y}^{nm} = (L + (-1)^{n+m}R) I_s^{nm}(-M, M_W; M, M_W)
\eea
in terms of a function
\bea
I_s^{nm}(M_1, M_2; M_3, M_4) \equiv \frac{1}{\sqrt{\pi R}} \int_{-\pi R}^{\pi R} dy 
\sin \left(\frac{m}{R}y \right) F_{M_1,M_2}^{(n)}(y)^* F_{M_3, M_4}^{(0)}(y)~(m \ge 1). 
\label{3.13}
\eea

As the matter of fact, 
$I_s^{nm}$ is not an independent function, and is related to $I_c^{nm}$. 
This is because in the non-zero KK mode sector $\gamma_y^{(m)}$ behaves as a would-be N-G boson 
to be ``eaten" by $\gamma_\mu^{(m)}$; Higgs-like mechanism is operative and 
the coupling of $\gamma_y^{(m)}$ should be equivalent to that of the longitudinal component 
of $\gamma_\mu^{(m)}$ (``equivalence theorem"). 
Or such relation may be attributed to the quantum mechanical supersymmetry, as is shown in the Appendix B. 
Anyway, we get a relation between $I_s^{nm}$ and $I_c^{nm}$, which means
\bea
V_{\gamma_y}^{nm} = i \frac{m_n-(-1)^{m+n}m_d}{\frac{m}{R}} V_{\gamma_\mu}^{nm}~(m \ge 1). 
\eea

The interaction terms of $h_\mu$ and $h$ are derived in a similar way, 
and we just give the result: 
\bea
&&\frac{g}{2} h_\mu^{(n)} \bar{d}^{(n)} \gamma^\mu V_{h_\mu}^{nm} d^{(0)}, \\
&&V_{h_\mu}^{nm} = (L + (-1)^{m+n}R) I_s(M, M_W; M, M_W)~(m \ge 1), \\
&&-\frac{g}{2} h^{(n)} \bar{d}^{(n)} V_h^{nm} d^{(0)}, \\
&&V_h^{nm} = \left\{
\begin{array}{c}
(L + (-1)^{m+n} R) I_c^{nm}(-M, M_W; M, M_W)~(m \ge 1) \\
(L + (-1)^{n} R) I^{n0}(-M, M_W; M, M_W)~(m=0). \\
\end{array}
\right.
\eea
Again a relation holds:
\bea
V_h^{nm} = -i \frac{m_n - (-1)^{m+n}m_d}{\frac{m}{R}} V_{h_\mu}^{nm}~(m \ge 1). 
\eea

In the case of interaction terms of $\hat{Z}_\mu$ and $\hat{\phi}^0$, 
there appears a transition between $\hat{\psi}_2$ and $\hat{\psi}_3$ and 
the vertex functions are described by new types of functions, $\tilde{I}_c$ and $\tilde{I}_s$.
Namely, the interaction term of $\hat{Z}_\mu$ is (for an arbitrary integer $m$)
\bea
&&-\frac{g}{2} \hat{Z}_\mu^{(m)} \bar{d}^{(n)} \gamma^\mu V_{Z_\mu}^{nm} d^{(0)}, \\
&&V_{Z_\mu}^{nm} = (L - (-1)^{m+n}R)\frac{1}{\sqrt{2}} 
\left( \tilde{I}_c^{nm}(M, M_W; M, M_W) + i \tilde{I}_s^{nm}(M, M_W; M, M_W) \right), 
\eea
where
\bea
\tilde{I}_c^{nm}(M_1, M_2; M_3, M_4) &\equiv& \frac{1}{\sqrt{\pi R}} 
\int_{-\pi R}^{\pi R} dy \cos \left(\frac{m}{R}y \right) F_{M_1, M_2}^{(n)}(-y)^* F_{M_3, M_4}^{(0)}(y), \\
\tilde{I}_s^{nm}(M_1, M_2; M_3, M_4) &\equiv& \frac{1}{\sqrt{\pi R}} 
\int_{-\pi R}^{\pi R} dy \sin \left(\frac{m}{R}y \right) F_{M_1, M_2}^{(n)}(-y)^* F_{M_3, M_4}^{(0)}(y). 
\eea
The interaction term of $\hat{\phi}^0$ reads as
\bea
&&-\frac{g}{2} \hat{\phi}^{0(m)} \bar{d}^{(n)} V_{\phi^0}^{nm} d^{(0)}, \nonumber \\
&&V_{\phi^0}^{nm} = (L - (-1)^{m+n}R)\frac{i}{\sqrt{2}} 
\left( \tilde{I}_c^{nm}(-M, M_W; M, M_W) + i \tilde{I}_s^{nm}(-M, M_W; M, M_W) \right). \nonumber \\
\eea
Some relations hold between $\tilde{I}_s^{nm}$ and $\tilde{I}_c^{nm}$ which mean, 
\bea
V_{\phi^0}^{nm} = i\frac{m_n + (-1)^{m+n}m_d}{\frac{m}{R} + M_Z}V_{Z_\mu}^{nm}. 
\eea

Finally, in the case of interaction terms of $\hat{W}_\mu^+$ and $\hat{\phi}^+$ 
the transition between $\hat{\psi}_1$ and $\hat{\psi}_2$ or $\hat{\psi}_3$ appears, 
which makes the vertex function a little complicated: 
for $n \ge 0$ and an arbitrary integer $m$, 
\bea
&&\frac{g}{2} \hat{W}_\mu^{+(m)} \bar{u}^{(n)} \gamma^\mu V_{W_\mu^+}^{nm} d^{(0)}, \\
&&V_{W_\mu^+}^{nm} = 
\left\{
\begin{array}{c}
\frac{1}{\sqrt{2}} \left( I_c^{nm}(M, 0; M, M_W) +i I_s^{nm}(M, 0; M, M_W) \right) (L + (-1)^{n+m}R) \\
+ \frac{1}{\sqrt{2}} \left( \tilde{I}_c^{nm}(M, 0; M, M_W) +i \tilde{I}_s^{nm}(M, 0; M, M_W) \right) (L - (-1)^{n+m}R)~(n \ge 1), \\
\left[I_c^{0m}(M, 0; M, M_W) +i  I_s^{0m}(M, 0; M, M_W) \right] L~(n=0), \\
\end{array}
\right. \nonumber \\
\\
&&-\frac{g}{2} \hat{\phi}^{+(m)} \bar{u}^{(n)} V_{\phi^+}^{nm} d^{(0)}, \\
&&V_{\phi^+}^{nm} = 
\left\{
\begin{array}{c}
\frac{1}{\sqrt{2}} \left( I_c^{nm}(-M, 0; M, M_W) +i I_s^{nm}(-M, 0; M, M_W) \right) (L + (-1)^{n+m}R) \\
-\frac{1}{\sqrt{2}} \left( \tilde{I}_c^{nm}(-M, 0; M, M_W) +i \tilde{I}_s^{nm}(-M, 0; M, M_W) \right) (L - (-1)^{n+m}R)~(n \ge 1), \\
\left[I_c^{0m}(-M, 0; M, M_W) +i  I_s^{0m}(-M, 0; M, M_W) \right] (-1)^m R~(n=0). \\
\end{array}
\right. \nonumber \\
\eea
Writing $V_{W_\mu^+}^{nm}$ as
\bea
V_{W_\mu^+}^{nm} = \alpha^{nm} L + \beta^{nm} R, 
\eea
$V_{\phi^+}^{nm}$ is written as (noting $\tilde{m}_0=0$), 
\bea
V_{\phi^+}^{nm} = -\frac{\tilde{m}_n \alpha^{nm} - m_d \beta^{nm}}{\frac{m}{R} +M_W} L 
-\frac{\tilde{m}_n \beta^{nm} - m_d \alpha^{nm}}{\frac{m}{R} +M_W} R. 
\eea

\begin{figure}[ht]
 \begin{center}
  \includegraphics[width=8cm]{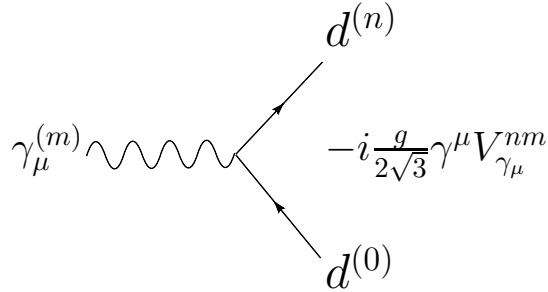}
 \end{center}
\caption{
The KK photon vertex between the down quark and KK fermion. 
} 
\label{photongaugeint}
\end{figure}

The Feynman rules for the $D$-dimensional gauge and Yukawa interactions of $\tilde{A}_\mu, \tilde{A}_y$ 
and $\hat{\psi}$ are readily read off from the results obtained above. 
For instance, the Feynman rule for the $\gamma_\mu^{(n)}~(n \ge 1)$ vertex is given by Fig. \ref{photongaugeint}. 

We also need the Feynman rule for the three point self-interactions of the photon $\gamma_\mu^{(0)}$ 
with charged gauge and Higgs bosons $\hat{W}_\mu^{\pm(n)}, \hat{\phi}^{\pm(n)}$ 
for the calculation of $g-2$. 
We skip all the detail of the derivation and just display the result: 
the relevant terms of $D$-dimensional lagrangian are
\bea
&&i e_D \left\{
(\partial_\mu \gamma_\nu^{(0)} - \partial_\nu \gamma^{(0)}_\mu ) \hat{W}^{+(n)\mu} \hat{W}^{-(n)\nu} 
- (\partial_\mu \hat{W}_\nu^{+(n)} - \partial_\nu \hat{W}^{+(n)}_\mu ) \gamma^{(0)\mu} \hat{W}^{-(n)\nu} 
\right. \nonumber \\
&& \left. 
+ (\partial_\mu \hat{W}_\nu^{-(n)} - \partial_\nu \hat{W}^{-(n)}_\mu ) \gamma^{(0)\mu} \hat{W}^{+(n)\nu} 
\right\} \nonumber \\
&&+ e_D \gamma^{(0)\mu} 
\left\{
\left( \frac{n}{R} + M_W \right) (\hat{W}_\mu^{+(n)} \hat{\phi}^{-(n)} + \hat{W}_\mu^{-(n)} \hat{\phi}^{+(n)}) 
+ i [(\partial_\mu \hat{\phi}^{+(n)}) \hat{\phi}^{-(n)} - (\partial_\mu \hat{\phi}^{-(n)}) \hat{\phi}^{+(n)} ]
\right\}, \nonumber \\
\eea
where $n$ can be an arbitrary integer. 
A nice thing here is that if we regard $\hat{W}_\mu^{+(n)}$ as $W_\mu^{+(0)}$, 
the lagrangian is just the same as that in the standard model, 
except that now $M_W$ is replaced by $\frac{n}{R} + M_W$. 
Thus to obtain the Feynman rule is straightforward. 
We get, {\it e.g.}, 
\begin{figure}[ht]
 \begin{center}
  \includegraphics[width=13cm]{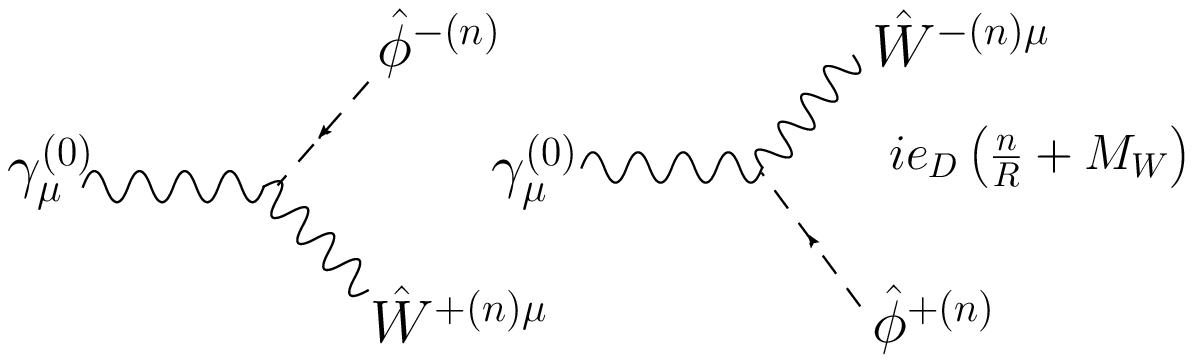}
 \end{center}
\end{figure}

\section{General formulae for the anomalous magnetic moment}
In this section, we derive general formulae for $A_\mu$ and $A_y$-exchange diagrams 
contributing to the anomalous magnetic moment, 
where $A_\mu, A_y$ denote generic $D$-dimensional gauge and scalar bosons, respectively. 

\subsection{$A_\mu$-exchange diagram}

\begin{figure}[ht]
 \begin{center}
  \includegraphics[width=7cm]{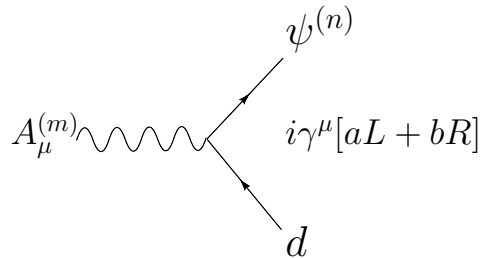}
 \end{center}
 \vspace*{-0.5cm}
\caption{
A generic gauge interaction vertex of the $D$-dimensional gauge boson $A_\mu^{(m)}$. 
} 
\label{gaugeint}
\end{figure}
We first derive a general formula for the $A_\mu^{(m)}$-exchange diagram due to the vertex in Fig. \ref{gaugeint}, 
with $A_\mu^{(m)}$ and $\psi^{(n)}$ being generic mass eigenstates of $A_\mu$ and fermion ($d$ or $u$), respectively, 
with masses $M_m$ and $m_n$. 
For simplicity, hereafter, we indicate $d^{(0)}$ just as $d$.

\begin{figure}[ht]
 \begin{center}
  \includegraphics[width=5cm]{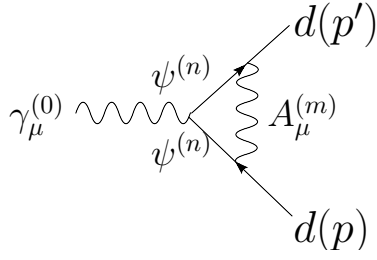} 
 \end{center}
 \vspace*{-0.5cm}
\caption{
The contribution of $A_\mu^{(m)}$-exchange to the anomalous magnetic moment. 
$p, p'$
are external momenta of $d$ quark. 
} 
\label{gaugeex}
\end{figure}

The diagram contributing to the anomalous magnetic moment is shown in Fig. \ref{gaugeex}. 
The $\gamma_\mu^{(0)}$ coupling at the tree level is modified 
due to the quantum correction into
\bea
-\frac{e_D}{3} \bar{d}(\gamma^\mu + \Gamma^\mu) d. 
\eea
Among a few term in $\bar{d} \Gamma^\mu d$, we are interested in the term 
proportional to $p^\mu + p'^\mu$ with a form factor $F_2(0)$.
\bea
\bar{d} \Gamma^\mu d \to \bar{d} \left[-\frac{1}{2m_d} (p^\mu + p'^\mu) F_2(0) \right] d. 
\label{gaugeex1}
\eea
It is the form factor $F_2(0)$ that gives the anomalous magnetic moment: 
$a=\frac{g-2}{2} = F_2(0)$. 

We obtain the contribution of Fig. \ref{gaugeex} to the form factor $F_2(0)$ as 
\bea
F_2^{A_\mu^{(m)}}(0) &=& -4i \frac{Q(\psi)}{Q(d)}
\int \frac{d^Dl}{(2\pi)^D} \int_0^1 dX X 
\times \nonumber \\ 
&& \frac{-\frac{a^*b+ab^*}{2} [4-DX] m_d m_n 
+ \frac{|a|^2+|b|^2}{2}(1 - X)[4-(D-2)X] m_d^2}{[l^2 + X(1-X)m_d^2 - X m_n^2 - (1 - X) M_m^2]^3} 
\label{g-2gauge}
\eea
where $Q(\psi)$ and $Q(d)$ are electric charges of $\psi^{(n)}$ and $d$ quark, respectively.

\subsection{$A_y$-exchange diagram}

\begin{figure}[ht]
 \begin{center}
  \includegraphics[width=6cm]{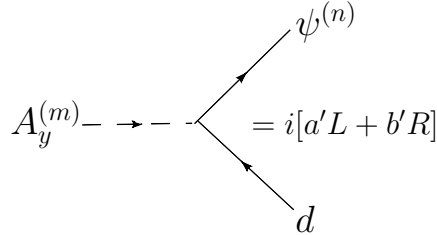} 
 \end{center}
 \vspace*{-0.5cm}
\caption{
A generic interaction vertex of the $D$-dimensional scalar $A_y^{(m)}$. 
} 
\label{Ayint}
\end{figure}

\begin{figure}[ht]
 \begin{center}
  \includegraphics[width=5cm]{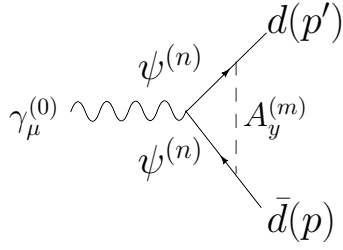}
 \end{center}
 \vspace*{-0.5cm}
\caption{
$A_y^{(m)}$-exchange diagram contributing to the anomalous magnetic moment. 
The external and internal momenta are defined in a similar way as in the $A_\mu^{(m)}$-exchange diagram. 
} 
\label{Ayex}
\end{figure}

Let us move to the calculation of $A_y^{(m)}$-exchange diagram due to the vertex shown in Fig. \ref{Ayint} 
where $A_y^{(m)}$ is a generic mass eigenstate of $A_y$ with a mass $M_m$. 
The diagram to evaluate is displayed in Fig. \ref{Ayex}. 
Similarly to the case of the $A_\mu^{(m)}$-exchange diagram, 
the one-loop correction is given by
\bea
F_2^{A_y^{(m)}}(0) = 4i \frac{Q(\psi)}{Q(d)}
\int \frac{d^Dl}{(2\pi)^D} \int_0^1 dX X^2  
\frac{\frac{a'^*b'+a'b'^*}{2} m_d m_n + \frac{|a'|^2+|b'|^2}{2} (1 - X) m_d^2}{[l^2 +X(1-X) m_d^2 - X m_n^2 - (1 - X) M_m^2]^3}. 
\label{g-2Ay}
\eea
\subsection{Diagrams due to the three point self-interaction}
\begin{figure}[ht]
 \begin{center}
  \includegraphics[width=5cm]{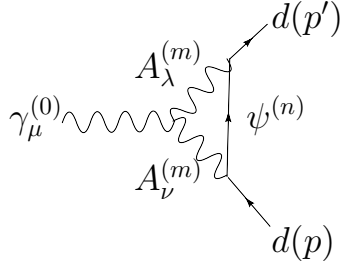}
 \end{center}
 \vspace*{-0.5cm}
\caption{
A diagram contributing to the anomalous magnetic moment due to a triple gauge boson vertex. 
} 
\label{3ptgauge}
\end{figure}

\begin{figure}[ht]
 \begin{center}
  \includegraphics[width=10cm]{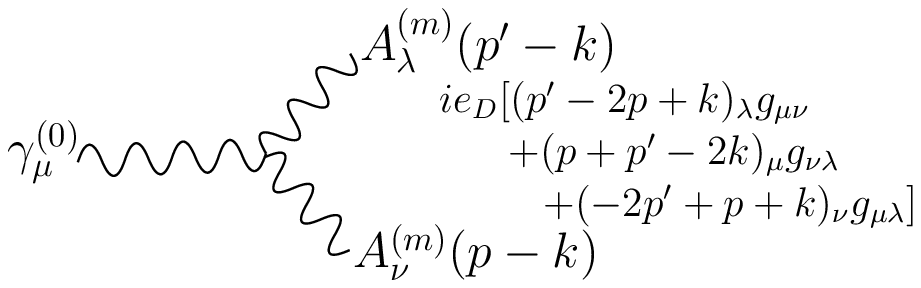}
 \end{center}
 \vspace*{-0.5cm}
\caption{
A triple gauge boson vertex. 
} 
\label{3ptvtx}
\end{figure}

We also need a general formula for the diagrams due to the three point self-interactions of 
$\gamma_\mu^{(0)}$ with $A_\mu^{(m)}$ and/or $A_y^{(m)}$. 
We first consider the contribution of a diagram (Fig. \ref{3ptgauge})
due to the vertex shown in Fig. \ref{gaugeint} and Fig. \ref{3ptvtx}
where the factor $e_D$ is assigned assuming $A_\mu^{(m)}$ is $\hat{W}_\mu^{-(m)}$, 
as is really the case. 
The contribution of Fig. \ref{3ptgauge} to $F_2(0)$ is given as
\bea
F_2(0)^{A_\mu^{(m)} A_\nu^{(m)}} &=& 
12i \int \frac{d^Dl}{(2\pi)^D} \int_0^1 dX X \times \nonumber \\
&& \frac{-\frac{a^*b + ab^*}{2}[4-D+(D-1)X]m_d m_n 
+ \frac{|a|^2+|b|^2}{2} X [5-D+(D-2)X] m_d^2}{[l^2 + X(1-X)m_d^2 -X M_m^2 -(1-X)m_n^2]^3}. 
\nonumber \\
\eea

\begin{figure}[ht]
 \begin{center}
  \includegraphics[width=10cm]{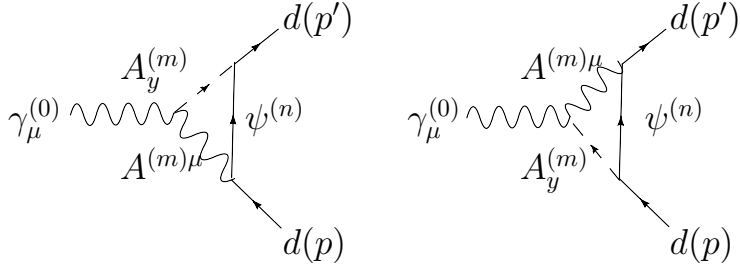}
 \end{center}
 \vspace*{-0.5cm}
\caption{
Diagrams contributing to the anomalous magnetic moment 
due to the $\gamma_\mu$-$A^\mu$-$A_y$ coupling. 
} 
\label{3ptmix}
\end{figure}

\begin{figure}[ht]
 \begin{center}
  \includegraphics[width=6cm]{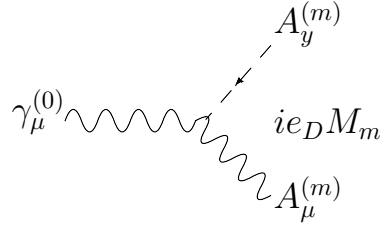}
 \end{center}
 \vspace*{-0.5cm}
\caption{
The 3-point vertex relevant for Fig. \ref{3ptmix}.
} 
\label{mixed}
\end{figure}

Next we consider the contribution of combined two diagrams shown in Fig. \ref{3ptmix}
due to the vertices in Fig. \ref{gaugeint}, Fig. \ref{Ayint} and additional one, Fig. \ref{mixed}. 
The contribution to $F_2(0)$ from Fig. \ref{3ptmix} is
\bea
F_2^{A_\mu^{(m)}A_y^{(m)}}(0) = -12i M_m \int \frac{d^Dl}{(2\pi)^D} \int_0^1 dX X 
\frac{\frac{1}{4}(b'^* a + a'^* b + b' a^* + a'b^*)X m_d}{[l^2 + X(1-X) m_d^2 - X M_m^2 - (1-X) m_n^2]^3}. 
\nonumber \\
\eea

\begin{figure}[ht]
 \begin{center}
  \includegraphics[width=5.5cm]{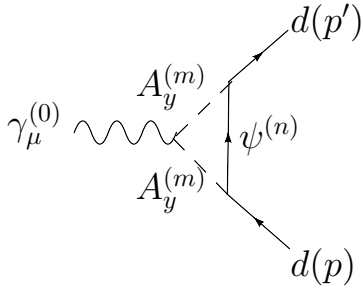}
 \end{center}
 \vspace*{-0.5cm}
\caption{
A diagram contributing to the anomalous magnetic moment 
due to the $\gamma_\mu$-$A_y$-$A_y$ coupling. 
} 
\label{gyy}
\end{figure}

\begin{figure}[ht]
 \begin{center}
  \includegraphics[width=8cm]{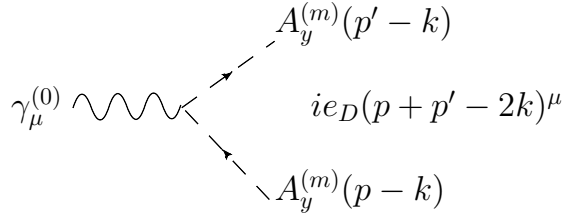}
 \end{center}
 \vspace*{-0.5cm}
\caption{
The 3-point vertex relevant for Fig. \ref{gyy}. 
} 
\label{3ptgyy}
\end{figure}

Finally, we consider the contribution of Fig. \ref{gyy}
due to another additional vertex of Fig. \ref{3ptgyy}. 
The contribution to $F_2(0)$ is
\bea
F_2^{A_y^{(m)}A_y^{(m)}}(0) &=& -12i \int \frac{d^Dl}{(2\pi)^D} \int_0^1 dX X \times \nonumber \\
&& \frac{
\frac{1}{2}(a'^* b' + a'b'^{*})(1-X) m_d m_n 
+ \frac{1}{2}(|a'|^2 + |b'|^2) X(1-X) m_d^2}{[l^2 + X(1-X) m_d^2 - X M_m^2 - (1-X) m_n^2]^3}. 
\eea

Having the general formulae, 
the contribution of each type of $A_\mu^{(m)}$ or $A_y^{(m)}$-exchange diagram is readily written down 
by use of the vertex functions derived in the previous section. 
For instance, the contribution of $\gamma_\mu^{(m)}$-exchange diagram $(m \ge 1)$ 
is obtained by setting (see (\ref{3.3}), (\ref{3.4}))
\bea
a = (-1)^{m+n} b = -\frac{g}{2\sqrt{3}} I_c^{nm}(M, M_W; M, M_W)
\eea
in (\ref{g-2gauge}): 
\bea
F_2^{\gamma_\mu^{(m \ne 0)}}(0) &=& -\frac{g^2}{3} i \sum_{m=1}^\infty \sum_{n=-\infty}^\infty 
\left[ I_c^{nm}(M, M_W; M, M_W) \right]^2 \int \frac{d^Dl}{(2\pi)^D} \int_0^1 dX X \times  \nonumber \\
&& \frac{-(-1)^{n+m} [4-DX] m_d m_n +(1-X) [4-(D-2)X] m_d^2}{[l^2 + X(1-X) m_d^2 -X m_n^2 -(1-X)(\frac{m}{R})^2]^3}. 
\eea
To get the other contributions is also straightforward and we do not write them down explicitly here. 
Even though in this way we can get the exact formulae for the contributions to the magnetic moment, 
to get the final analytic results is a hard task, since the vertex functions like $I_c(M, M_W; M, M_W)$ 
are rather complicated and $m_n$ to describe the function cannot be given analytically, 
unless some approximation is applied or some extreme cases are considered.

\section{The cancellation of divergences}

We are now ready to focus on the main issue of this paper, 
{\em i.e.} the cancellation mechanism of UV divergence in the contribution to $g-2$. 
As discussed in the introduction, 
the anomalous magnetic moment in the present model is expected to be finite 
relying on the operator analysis, similarly to the case of toy model, 
$(D+1)$ dimensional QED \cite{ALM}. 
In Ref. \cite{ALM}, the cancellation seems to take place 
between photon-exchange and Higgs-exchange diagrams. 
In the present model, however, these contributions behave differently. 
For instance, the Yukawa coupling of Higgs to $d$ quark is suppressed compared to the gauge coupling. 
We will find that in the present realistic model, 
the cancellation takes place between the contributions of the partners $(A_\mu^{(m)}, A_y^{(m)})$ 
having the same quantum number, such as $(\hat{W}_\mu^{+(m)}, \hat{\phi}^{+(m)})$, 
among which the Higgs(-like) mechanism is at work. 
It is interesting to realize that they also form ``super-partners" of the quantum mechanical SUSY 
present in the higher dimensional gauge theories \cite{LNSS}.

As has been discussed in the end of previous section, 
the exact formulae for the contributions of each diagram are not easy to handle. 
Since the cancellation should occur for the arbitrary bulk mass $M$, 
being supported by the operator analysis, 
and our main purpose is to confirm the cancellation mechanism, 
we simplify the analysis by considering the case of small $M$ in this section. 
We will retain only the term up to ${\cal O}(M)$ in the contribution of each diagram. 
As the matter of fact, 
it turns out that the term linear in $M$ in each contribution vanishes. 
This comes from the fact that the functions $I_c^{nm}$, {\em etc.}, 
appearing in the fermion vertices satisfy a relation, as is shown in Appendix A: 
\bea
I_c^{nm}(-M, M_W; -M, M_W) = (-1)^{n+m} I_c^{nm}(M, M_W; M, M_W). 
\eea
Since such functions appear twice in each diagram, 
we conclude the contribution of each diagram is an even function of $M$. 
Hence the term linear in $M$ actually disappears. 
We thus take the limit of $M \to 0$. 

In the limit $M \to 0$, $m_n$ reduces to $\frac{n}{R}+M_W$ and the functions $I_c^{nm}$ etc. 
are greatly simplified, reflecting the (partial) recovery of momentum conservation 
along the extra dimension: 
\bea
&&I_c^{nm}(0, M_2; 0, M_4) = \frac{1}{2\sqrt{\pi R}} (\delta_{n,m} + \delta_{-n,m})~(m \ne 0), \\
&&I_c^{n0}(0, M_2; 0, M_4) = \frac{1}{\sqrt{2\pi R}} \delta_{n,0}, \\
&&I_s^{nm}(0, M_2; 0, M_4) = \frac{i}{2\sqrt{\pi R}} (\delta_{n,m} - \delta_{-n,m})~(m \ne 0), \\
&&\tilde{I}_c^{nm}(0, M_2; 0, M_4) = \frac{1}{2\sqrt{\pi R}} (\delta_{n,m} + \delta_{-n,m})~(m \ne 0), \\
&&\tilde{I}_s^{nm}(0, M_2; 0, M_4) = -\frac{i}{2\sqrt{\pi R}} (\delta_{n,m} - \delta_{-n,m})~(m \ne 0).
\eea
Then the vertex functions such as $V_{\gamma_\mu}^{nm}$ in this limit 
also take simple forms. 
For instance, $V_{\gamma_\mu}^{nm}=\frac{1}{2\sqrt{\pi R}}(\delta_{n,m} + \delta_{-n,m})~(m \ge 1)$. 
Having the Feynman rules for the interaction vertices of fermions, 
we readily get the contribution to $g-2$, {\em i.e.} $F_2(0)$, 
from each one-loop diagram, by use of the general formulae derived in section 4. 
We display the results below, dividing them into the sectors of 
charged current, neutral current, etc. 

\subsection{The charged current sector}

We first display the contribution of $\hat{W}_\mu^+$-exchange 
diagram (obtained with $A_\mu^{(m)}$ being replaced by $\hat{W}_\mu^{+(\pm n)}$ in Fig. \ref{gaugeex}),
which we denote by $F_2^{(W)}(0)$: 
\bea
F_2^{(W)}(0) &=& -4i (-2) \left( \frac{g_D}{2} \right)^2 \int \frac{d^Dl}{(2\pi)^D} 
\int_0^1dX X \nonumber \\
&& \times \sum_{n=-\infty}^\infty 
\frac{(4-DX)\frac{n}{R}M_W +(1-X) [4-(D-2)X]M_W^2}{[l^2-(\frac{n}{R}+(1-X)M_W)^2]^3}
\label{F2W}
\eea
where the factor $(-2)$ comes from $Q(u)/Q(d)$ and 
the facts $m_d = M_W$, $\tilde{m}_n =\frac{n}{R}$, $M_{\hat{W}_\mu^{+(n)}} = \frac{n}{R} + M_W$ 
(in the limit $M \to 0$) have been used. 
A wisdom to treat the UV divergence is to invoke Poisson resummation 
and extract the ``zero-winding" sector. 
Here, however, we just take the limit of ``de-compactification", $R\to \infty$, 
as the zero-winding sector is easily known to correspond to the limit. 
In this limit $\frac{n}{R}$ may be replaced by the continuous extra space momentum $p_y$. 
Thus, taking the limit $R \to \infty$, replacing $\frac{n}{R}$ by $p_y$ 
and performing a shift of the momentum, $p_y +(1-X)M_W \to p_y$, 
we get the divergent part $F_2(0)_{{\rm div.}}$ of the contribution, 
\bea
F_2^{(W)}(0)_{{\rm div.}} = i \frac{2\pi}{3} R g_D^2 \int \frac{d^Dl d p_y}{(2\pi)^{D+1}} \frac{M_W^2}{[l^2-p_y^2]^3}. 
\label{Wexdiv}
\eea

We next display the contribution of the exchange of the partner of $W_\mu^+$, 
{\em i.e.} $\hat{\phi}^+$, whose diagram is obtained with $A_y^{(m)}$ being replaced by $\hat{\phi}^{+(\pm n)}$ in Fig. \ref{Ayex}. 
\bea
F_2^{(\phi^+)}(0) = 4i (-2) \left(\frac{g_D}{2} \right)^2 \int \frac{d^Dl}{(2\pi)^D} \int_0^1 dX X 
\sum_{n=-\infty}^\infty \frac{-X \frac{n}{R} M_W + X(1-X)M_W^2}{[l^2-(\frac{n}{R} + (1-X) M_W)^2]^3}. 
\label{F2phi}
\eea
Taking the same step as we took above, 
we obtain the divergent part of the contribution, 
\bea
F_2^{(\phi^+)}(0)_{{\rm div.}} =-i \frac{2\pi}{3} Rg_D^2 \int \frac{d^Dl dp_y}{(2\pi)^{D+1}} 
\frac{M_W^2}{[l^2-p_y^2]^3}. 
\label{phiexdiv}
\eea
We realize , as we anticipated, the divergence exactly cancel out 
between the contributions of ``partners" $(\hat{W}_\nu^{+(n)}, \hat{\phi}^{+(n)})$ 
shown in (\ref{Wexdiv}) and (\ref{phiexdiv}), 
though the original forms of (\ref{F2W}) and (\ref{F2phi}) are quite different. 
Let us note that the Higgs(-like) mechanism is operative 
both in the non-zero KK mode sector $(n \ne 0)$ and the zero-mode sector $(n=0)$ 
for the charged gauge-Higgs bosons, being triggered by the KK mass $\frac{n}{R}$ for $n \ne 0$ 
and by the VEV of $A_y$ for $n=0$ ({\em i.e.} the ordinary Higgs mechanism).  

In addition to these diagrams, 
we have to evaluate the contributions due to the three point self-interaction vertices, 
whose diagrams are obtained with $A_\mu^{(m)}$ and $A_y^{(m)}$ being replaced 
by $\hat{W}_\mu^{+(\pm n)}$ and $\hat{\phi}^{+(\pm n)}$ 
in Figs. \ref{3ptgauge}, \ref{3ptmix}, and \ref{gyy}. 
  %
  %
  %
  %
The contributions of each diagram are
\bea
F_2^{(WW)}(0) &=& 3ig_D^2 \int \frac{d^Dl}{(2\pi)^D} \int_0^1 dX X \times \nonumber \\
&& \sum_{n=-\infty}^\infty 
\frac{-[D-4-(D-1)X] \frac{n}{R} M_W + X[5-D+(D-2)X]M_W^2}{[l^2-(\frac{n}{R} + X M_W)^2]^3}, \\
F_2^{(W \phi^+)}(0) &=& 3ig_D^2 \int \frac{d^Dl}{(2\pi)^D} \int_0^1 dX X \sum_{n=-\infty}^\infty 
\frac{X \frac{n}{R} M_W + X M_W^2}{[l^2-(\frac{n}{R} + X M_W)^2]^3}, \\
F_2^{(\phi^+ \phi^+)}(0) &=& -3ig_D^2 \int \frac{d^Dl}{(2\pi)^D} \int_0^1 dX X \sum_{n=-\infty}^\infty 
\frac{-(1-X) \frac{n}{R} M_W + X (1-X) M_W^2}{[l^2-(\frac{n}{R} + X M_W)^2]^3}. 
\eea
It is straightforward to show that 
the divergent part just cancel out; 
\bea
F_2^{(WW)}(0) &\stackrel{R \to \infty}{\longrightarrow}& 6\pi i R g_D^2 \int \frac{d^Dl dp_y}{(2\pi)^{D+1}} 
\int_0^1 dX X \frac{X(1-X)M_W^2}{(l^2-p_y^2)^3}, \\
F_2^{(W \phi^+)}(0) &\stackrel{R \to \infty}{\longrightarrow}& 6\pi i R g_D^2 \int \frac{d^Dl dp_y}{(2\pi)^{D+1}} 
\int_0^1 dX X \frac{X(1-X)M_W^2}{(l^2-p_y^2)^3}, \\
F_2^{(\phi^+ \phi^+)}(0) &\stackrel{R \to \infty}{\longrightarrow}& -12\pi i R g_D^2 \int \frac{d^Dl dp_y}{(2\pi)^{D+1}} 
\int_0^1 dX X \frac{X(1-X)M_W^2}{(l^2-p_y^2)^3}, 
\eea
and 
\bea
\left( F_2^{(WW)}(0) + F_2^{(W \phi^+)}(0) + F_2^{\phi^+ \phi^+}(0) \right)_{{\rm div.}} = 0. 
\eea

\subsection{The neutral current sector}
Taking a similar step to the case of the charged current sector, 
the contribution due to $\hat{Z}_\mu$-exchange diagram 
is known to be given as
\bea
F_2^{(Z)}(0) &=& -4i \left(\frac{g_D}{2} \right)^2 \int \frac{d^Dl}{(2\pi)^D} \int_0^1 dX X \times  \nonumber \\
&& 
\sum_{n=-\infty}^\infty \frac{(4-DX)(\frac{n}{R}+M_W)M_W +(1-X) [4 - (D-2)X] M_W^2}{[l^2-(\frac{n}{R}+(2-X)M_W)^2]^3},
\eea
where $m_n = \frac{n}{R}+M_W, M_{\hat{Z}_\mu^{(n)}} = \frac{n}{R} + 2M_W$ has been used.  

The $R\to \infty$ limit and the shift $p_y +(2-X) M_W \to p_y$ gives
\bea
F_2^{(Z)}(0)_{{\rm div.}} = -i \frac{\pi}{3}R g_D^2 \int \frac{d^Dl dp_y}{(2\pi)^{D+1}} \frac{M_W^2}{(l^2-p_y^2)^3}. 
\eea

The contribution due to $\hat{\phi}^0$-exchange diagram 
is given as
\bea
F_2^{(\phi^0)}(0) &=& 4i \left(\frac{g_D}{2} \right)^2 \int \frac{d^Dl}{(2\pi)^D} \int_0^1 dX X \times  \nonumber \\
&& 
\sum_{n=-\infty}^\infty \frac{-X (\frac{n}{R}+M_W) M_W + X(1-X) M_W^2}{[l^2-(\frac{n}{R}+(2-X)M_W)^2]^3}. 
\eea
The divergent part reads as
\bea
F_2^{(\phi^0)}(0)_{{\rm div.}} = i \frac{\pi}{3}R g_D^2 \int \frac{d^Dl dp_y}{(2\pi)^{D+1}} \frac{M_W^2}{(l^2-p_y^2)^3}. 
\eea
We thus confirm the cancellation of the divergence,
\bea
F_2^{(Z)}(0)_{{\rm div.}} + F_2^{(\phi^0)}(0)_{{\rm div.}} = 0.  
\eea

\subsection{The photon sector}
In this subsection, we discuss the contributions of photon ($\gamma_\mu^{(n)}$)-exchange 
and the exchange of $\gamma_y^{(n)}$, the partner of $\gamma_\mu^{(n)}$ for $n \ge 1$. 
In clear contrast to the previous cases, 
the Higgs mechanism does not exist for the zero mode sector, $n=0$, 
and also $\gamma_y^{(0)}$ is absent due to the orbifolding. 
This suggests that the contribution of $n=0$ sector differs from 
what we obtain by formally setting $n=0$ in the formula valid for $n \ne 0$. 
In fact, the contribution of the $\gamma_\mu^{(n)}$-exchange ($n \ge 0$)
is given by 
\bea
F_2^{(\gamma_\mu)}(0) &=& -4i \left(-\frac{e_D}{3\sqrt{2}} \right)^2 \int \frac{d^Dl}{(2\pi)^D} 
\int_0^1 dX X \times \nonumber \\
&& \sum_{n=-\infty}^\infty \frac{-(4-DX)(\frac{n}{R}+M_W)M_W + (1-X)[4-(D-2)X] M_W^2}{[l^2-(\frac{n}{R} + X M_W)^2]^3} 
\nonumber \\
&&-4i \left(-\frac{e_D}{3\sqrt{2}} \right)^2 \int \frac{d^Dl}{(2\pi)^D} \int_0^1 dXX^2 \frac{[(D-2)X-2]M_W^2}{(l^2 - X^2 M_W^2)^3}. 
\label{ptng-2}
\eea
The second term is to adjust the discrepancy mentioned above. 
The divergent part reads as 
\bea
F_2^{(\gamma_\mu)}(0)_{{\rm div.}} &=& -i \frac{2\pi}{27} R e_D^2 \int \frac{d^Dl dp_y}{(2\pi)^{D+1}} \frac{M_W^2}{(l^2-p_y^2)^3} 
\nonumber \\ 
&& -i \frac{2}{9} e_D^2 \int \frac{d^Dl}{(2\pi)^D} \int_0^1 dX X^2 \frac{[(D-2)X-2] M_W^2}{(l^2-X^2M_W^2)^3}. 
\label{ptndiv}
\eea
Similarly, the contribution of the $\gamma_y^{(n)}$-exchange ($n \ge 1$) 
is given by 
\bea
F_2^{(\gamma_y)}(0) &=& 4i \left(\frac{e_D}{3\sqrt{2}} \right)^2 \int \frac{d^Dl}{(2\pi)^D} 
\int_0^1 dX X \times \nonumber \\
&& \sum_{n=-\infty}^\infty \frac{X(\frac{n}{R}+M_W)M_W + X(1-X) M_W^2}{[l^2-(\frac{n}{R} + X M_W)^2]^3} 
\nonumber \\
&&-4i \left(\frac{e_D}{3\sqrt{2}} \right)^2 \int \frac{d^Dl}{(2\pi)^D} \int_0^1 dXX^2 \frac{(2-X) M_W^2}{(l^2 - X^2 M_W^2)^3}, 
\label{Ayg-2}
\eea
whose divergent part reads as 
\bea
F_2^{(\gamma_y)}(0)_{{\rm div.}} &=& i \frac{2\pi}{27} R e_D^2 \int \frac{d^Dl dp_y}{(2\pi)^{D+1}} \frac{M_W^2}{(l^2-p_y^2)^3} 
\nonumber \\ 
&& -i \frac{2}{9} e_D^2 \int \frac{d^Dl}{(2\pi)^D} \int_0^1 dX X^2 \frac{(2-X) M_W^2}{(l^2-X^2M_W^2)^3}. 
\label{Aydiv}
\eea
We thus realize that although the cancellation of divergence is ``almost" complete, 
there remains a ``partial" $D$-dimensional divergence originated from the $n=0$ sector, 
\bea
F_2^{(\gamma_\mu)}(0)_{{\rm div.}} + F_2^{(\gamma_y)}(0)_{{\rm div.}} 
= -i \frac{2}{9} (D-3)e_D^2 \int \frac{d^Dl}{(2\pi)^D} \int_0^1 dX X^3 \frac{M_W^2}{(l^2-X^2 M_W^2)^3}. 
\label{zeromodediv}
\eea

\subsection{The Higgs sector}

Finally, in this subsection we discuss the contribution of the exchange 
of the partner of the Higgs $h_\mu^{(n)}$, and the contribution of the Higgs ($h^{(n)}$)-exchange. 
The situation concerning the UV divergence is similar to that in the photon sector, 
and we just summarize the result below. 

The contribution of the $h_\mu^{(n)}$-exchange ($n \ge 1$)
is given by 
\bea
F_2^{(h_\mu)}(0) &=& -4i \left(\frac{g_D}{2\sqrt{2}} \right)^2 \int \frac{d^Dl}{(2\pi)^D} 
\int_0^1 dX X \times \nonumber \\
&& \sum_{n=-\infty}^\infty \frac{-(4-DX)(\frac{n}{R}+M_W)M_W + (1-X)[4-(D-2)X] M_W^2}{[l^2-(\frac{n}{R} + X M_W)^2]^3} 
\nonumber \\
&&+4i \left(\frac{g_D}{2\sqrt{2}} \right)^2 \int \frac{d^Dl}{(2\pi)^D} \int_0^1 dX X^2 \frac{[(D-2)X-2]M_W^2}{(l^2 - X^2 M_W^2)^3}, 
\label{hmug-2}
\eea
whose divergent part reads as 
\bea
F_2^{(h_\mu)}(0)_{{\rm div.}} &=& -i \frac{\pi}{6} R g_D^2 \int \frac{d^Dl dp_y}{(2\pi)^{D+1}} \frac{M_W^2}{(l^2-p_y^2)^3} 
\nonumber \\ 
&& +i \frac{1}{2} g_D^2 \int \frac{d^Dl}{(2\pi)^D} \int_0^1 dX X^2 \frac{[(D-2)X-2] M_W^2}{(l^2-X^2M_W^2)^3}. 
\label{hmudiv}
\eea

The contribution of the $h^{(n)}$-exchange ($n \ge 0$) 
is given by 
\bea
F_2^{(h)}(0) &=& 4i \left(\frac{g_D}{2\sqrt{2}} \right)^2 \int \frac{d^Dl}{(2\pi)^D} 
\int_0^1 dX X \times \nonumber \\
&& \sum_{n=-\infty}^\infty \frac{X(\frac{n}{R}+M_W) M_W + X(1-X) M_W^2}{[l^2-(\frac{n}{R} + X M_W)^2]^3} 
\nonumber \\
&&+4i \left(\frac{g_D}{2\sqrt{2}} \right)^2 \int \frac{d^Dl}{(2\pi)^D} \int_0^1 dX X^2 \frac{(2-X) M_W^2}{(l^2 - X^2 M_W^2)^3}, 
\label{hg-2}
\eea
whose divergent part reads as 
\bea
F_2^{(h)}(0)_{{\rm div.}} &=& i \frac{\pi}{6} R g_D^2 \int \frac{d^Dl dp_y}{(2\pi)^{D+1}} \frac{M_W^2}{(l^2-p_y^2)^3} 
\nonumber \\ 
&& +i \frac{1}{2} g_D^2 \int \frac{d^Dl}{(2\pi)^D} \int_0^1 dX X^2 \frac{(2-X) M_W^2}{(l^2-X^2M_W^2)^3}. 
\label{Aydiv}
\eea
Again, the cancellation of UV divergence turns out to be not complete, 
\bea
F_2^{(h_\mu)}(0)_{{\rm div.}} + F_2^{(h)}(0)_{{\rm div.}} 
= i \frac{1}{2} (D-3)g_D^2 \int \frac{d^Dl}{(2\pi)^D} \int_0^1 dX X^3 \frac{M_W^2}{(l^2-X^2M_W^2)^3}. 
\label{zeromodehdiv}
\eea

To summarize this section, 
we have found the cancellation mechanism of UV divergence 
between the contributions of the pairs of $D$-dimensional vector bosons and scalars 
$(A_\mu, A_y)$, which form ``partners" of Higgs(-like) mechanism and at the same time 
the partners of quantum mechanical SUSY. 
The cancellation is complete for the charged current and neutral current sectors, 
$(\hat{W}_\mu^{+(n)}, \hat{\phi}^{+(n)})$ and $(\hat{Z}_\mu^{(n)}, \hat{\phi}^{0(n)})$, 
while it is incomplete for the photon and Higgs sectors, 
$(\gamma_\mu^{(n)}, \gamma_y^{(n)})$ and $(h_\mu^{(n)}, h^{(n)})$. 
The remaining ``partial" and lower dimensional ($D$-dimensional) divergence 
({\rm i.e.} the sum of Eqs. (\ref{zeromodediv}) and (\ref{zeromodehdiv}), by use of $e_D^2 =\frac{3}{4} g_D^2$) 
\bea
F_2(0)_{{\rm div.}} = i \frac{1}{3}(D-3)g_D^2 \int \frac{d^Dl}{(2\pi)^D} 
\int_0^1 dX X^3 \frac{M_W^2}{(l^2 -X^2 M_W^2)^3}, 
\label{remaindiv}
\eea
seems to be attributed to the absence of Higgs mechanism for $n=0$ in the photon and Higgs sectors. 
As the matter of fact, however, Eq.(\ref{remaindiv}) is finite, so is the anomalous magnetic moment, 
for 5 or 6 dimensional ($D=4~{\rm or}~5$) space-time. 
Let us note the $g-2$ is divergent already at 6 dimensional space-time 
in other types of higher dimensional gauge theories, 
{\em e.g.} in the scenario of universal extra dimension.

\section{Conclusions}
In this paper, 
we have discussed the finiteness of the anomalous magnetic moment of fermion 
in a realistic model of gauge-Higgs unification. 
Our main purpose is to clarify the cancellation mechanism of UV divergences 
in various contributing Feynman diagrams. 
Our expectation that the anomalous moment should be finite and calculable, 
in spite of the fact that higher dimensional gauge theories are argued to be non-renormalizable, 
is based on an operator analysis: 
an operator corresponding to the dimension six gauge invariant operator 
describing the anomalous moment in the standard model 
is forbidden because of the higher dimensional gauge symmetry 
present in the gauge-Higgs unification and the on-shell condition for the fermion, 
as was discussed in the introduction.   

In our previous paper, 
we adopted a toy model, i.e. higher dimensional QED compactified on $S^{1}$, 
where $D$-dimensional gauge and scalar fields $(A_{\mu}, A_{y})$ were identified with photon and Higgs fields, respectively, 
and showed by explicit calculation that the anomalous moment is in fact finite 
for arbitrary space-time dimensions \cite{ALM}. 
Although the result is quite remarkable, 
this toy model is not realistic and has a few drawbacks: 
the gauge group is too small to incorporate the $SU(2)$ doublet Higgs 
and the masses of light fermions or their small Yukawa couplings cannot be taken into account. 
It cannot reproduce the famous result of Schwinger on the magnetic moment \cite{Schwinger}, 
again due to the unrealistic Yukawa coupling. 
 
These unsatisfactory points are overcomed in this paper 
by considering a realistic $D+1$ dimensional ($D = 4,5,$ etc.) $SU(3)$ gauge-Higgs unification model 
compactified on an orbifold $S^{1}/Z_{2}$, with matter fermions of $SU(3)$ triplet. 
The small Yukawa coupling is achieved by introducing $Z_{2}$-odd bulk mass $M$ for the fermions.

After deriving various general formulae which are valid for arbitrary $M$ 
and are used to obtain the anomalous moment, 
we have discussed the cancellation mechanism of the UV divergences 
for the simplified case of small $M$. 
Since our operator analysis concerning the finiteness strongly 
depends on the higher dimensional gauge symmetry, 
it will be natural to expect that the cancellation of UV divergence is realized 
between the contributions of the pair of $D$-dimensional gauge and scalar bosons 
$(A_{\mu}, A_{y})$ with the same quantum number. $(A_{\mu}, A_{y})$, at least for non-zero KK modes, 
play the roles as the ``partners" in the Higgs-like mechanism 
to form massive gauge bosons (from $D$-dimensional point of view), 
which is the manifestation of the higher dimensional gauge symmetry. 
We also would like to point out that the pair $(A_{\mu}, A_{y})$ is known to behave 
as a multiplet of quantum mechanical SUSY \cite{LNSS}, 
so the cancellation may be regarded as the consequence of the supersymmetry. 
Let us note that in this model photon and Higgs have different quantum numbers and 
the divergence cancellation does not take place between the contributions of these particles, 
in clear contrast to the case of our previous paper \cite{ALM}.  

We have confirmed these expectations by explicit calculation of Feynman diagrams. 
Concerning the contributions of charged and neutral currents 
due to the pairs $(\hat{W}_{\mu}^{\pm (n)}, \hat{\phi}^{\pm (n)})$ and $(\hat{Z}_{\mu}^{(n)}, \hat{\phi}^{0(n)})$, 
the zero-mode sector also has ordinary Higgs mechanism. 
Thus the UV divergence has been shown to be completely cancelled 
between the contributions of each partners. 
On the other hand, 
concerning the remaining contributions of photon and Higgs sector, 
$(\gamma_{\mu}^{(n)}, \gamma_{y}^{(n)})$ and $(h_{\mu}^{(n)}, h^{(n)})$, 
the cancellation is not complete and there remains a UV divergence (but of lower degree) 
which originates from the zero modes. 
Let us recall that for these sectors the zero modes do not have ordinary Higgs mechanism and 
one member of each partners is missing due to the orbifolding. 
So far, we do not have any good reasoning why the cancellation mechanism does not perfectly work, 
while the operator analysis seems to be valid. 
This issue remains to be settled. 
Nevertheless, we would like to point out 
that we have completely finite anomalous moments for 5 and 6 dimensional space-time, 
although the anomalous moment is divergent for the case of 6 dimensions 
in other higher dimensional gauge theories such as ``universal extra dimension". 

In our previous paper, 
we could not reproduce the Schwinger's result in ordinary QED, 
since the contribution of the Higgs-exchange was comparable to that of photon-exchange 
as the Yukawa coupling was of the order of the gauge coupling. 
This drawback is overcomed in the present model 
by introducing the bulk mass which causes the localization of Weyl fermions 
with different chiralities at two different fixed points. 
We will report in the forthcoming paper \cite{ALM2} 
that the Schwinger's result is indeed reproduced. 
In the paper, we will also discuss the constraint on the compactification scale 
by comparing our prediction on the anomalous magnetic moment 
including the contributions of non-zero KK modes 
with the allowed deviation of the data from the prediction of the standard model.

\subsection*{Acknowledgment}

The work of the authors was supported 
in part by the Grant-in-Aid for Scientific Research 
of the Ministry of Education, Science and Culture, No.18204024 and No. 20025005.  

\begin{appendix}

\setcounter{equation}{0}
\section{The derivation of the property \\$I_c^{nm}(-M, M_W; -M, M_W) = (-1)^{n+m} I_c^{nm}(M, M_W; M, M_W)$}
In this appendix, the relation between the vertex functions 
due to the fermions with different chiralities is discussed. 
We take a typical example of $I_c^{nm}(M, M_W; M, M_W)$ and 
$I_c^{nm}(-M, M_W; -M, M_W)$ defined in (\ref{3.6}) and (\ref{3.7}) 
to show the property \\$I_c^{nm}(-M, M_W; -M, M_W) = (-1)^{n+m} I_c^{nm}(M, M_W; M, M_W)$. 
Since the interchange of the chirality $L \leftrightarrow R$ corresponds 
to the sign flip of the bulk mass $M \to -M$, therefore the exchange of the fixed points 
$y \leftrightarrow y-\pi R$, we expect that we can obtain $f_{d_R}$ by shifting $f_{d_L}$ by $\pi R$ 
with possible phase change. 
As the shift does not mix the real and imaginary parts, 
we expect that such property holds in each part of even and odd functions of $y$. 

Let us start with left-handed mode function,
\bea
f_{d_L}^{(n)}(y) &=& F_{M, M_W}^{(n)}(y) \nonumber \\
&=& e^{iM_W y} C^{(n)} 
\left[
\sqrt{\cos(\varphi_{n} - \alpha_{n})} \cos(\sqrt{m_{n}^2-M^2}|y| + \alpha_{n}) 
\right. \nonumber \\
&& \left. -\varepsilon(n) i \sqrt{\cos(\varphi_{n} +\alpha_{n})} \sin(\sqrt{m_{n}^2-M^2}y)
\right]  \nonumber \\
&=&
C^{(n)} 
\left[
\cos(M_W y) \sqrt{\cos(\varphi_{n} - \alpha_{n})} \cos(\sqrt{m_{n}^2-M^2}|y| + \alpha_{n}) 
\right. \nonumber \\
&& \left. + \varepsilon(n) \sin(M_W y) \sqrt{\cos(\varphi_{n} + \alpha_{n})} \sin(\sqrt{m_{n}^2-M^2}y) 
\right. \nonumber \\
&& \left. +i \left(
-\varepsilon(n) \cos(M_W y) \sqrt{\cos(\varphi_{n} + \alpha_{n})} \sin(\sqrt{m_{n}^2-M^2}y) 
\right. \right. \nonumber \\
&& \left. \left. + \sin(M_W y) \sqrt{\cos(\varphi_{n} - \alpha_{n})} \cos(\sqrt{m_{n}^2-M^2}|y| + \alpha_{n}) 
\right) 
\right], 
\eea
where $\varepsilon(n)=1$ for $n \ge 0$, $-1$ for $n < 0$.
 
Concentrating on the real part of $f_{d_L}^{(n)}/C^{(n)}$ ($C^{(n)}$ is invariant under $M \to -M$), 
and making a shift $y \to y-\pi R$, we find
\bea
&&[f_{d_L}^{(n)}(y-\pi R)/C^{(n)}]_{{\rm real}} \nonumber \\
&=& 
\cos(M_W (y-\pi R)) \sqrt{\cos(\varphi_{n} - \alpha_{n})} 
\cos(\sqrt{m_{n}^2-M^2}(\pi R -y) + \alpha_{n}) \nonumber \\
&&+ \varepsilon(n) \sin(M_W (y-\pi R)) \sqrt{\cos(\varphi_{n} + \alpha_{n})} \sin(\sqrt{m_{n}^2-M^2}(y-\pi R)) \nonumber \\
&=& \cos(M_W y) 
\left[
\left(
\cos(M_W \pi R) \sqrt{\cos(\varphi_{n} - \alpha_{n})} \cos \varphi_{n} \right. \right. \nonumber \\
&& \left. \left. + \varepsilon(n) 
\sin(M_W \pi R) \sqrt{\cos(\varphi_{n} + \alpha_{n})} \sin (\varphi_{n} - \alpha_{n}) 
\right) \cos(\sqrt{m_{n}^2-M^2}y-\alpha_{n}) \right. \nonumber \\
&& \left. + \left(
\cos(M_W \pi R) \sqrt{\cos(\varphi_{n} - \alpha_{n})} \sin \varphi_{n} \right. \right. \nonumber \\
&& \left. \left. -\varepsilon(n) 
\sin(M_W \pi R) \sqrt{\cos(\varphi_{n} + \alpha_{n})} \cos (\varphi_{n} - \alpha_{n}) 
\right) \sin(\sqrt{m_{n}^2-M^2}y-\alpha_{n})
\right] \nonumber \\
&& + \sin(M_W y) 
\left[
\left(
\sin(M_W \pi R) \sqrt{\cos(\varphi_{n} - \alpha_{n})} \sin (\varphi_{n}+ \alpha_{n}) \right. \right. \nonumber \\
&& \left. \left. +\varepsilon(n) 
\cos(M_W \pi R) \sqrt{\cos(\varphi_{n} + \alpha_{n})} \cos \varphi_{n}  
\right) \sin(\sqrt{m_{n}^2-M^2}y) \right. \nonumber \\
&& \left. + \left(
\sin(M_W \pi R) \sqrt{\cos(\varphi_{n} - \alpha_{n})} \cos (\varphi_{n} + \alpha_{n}) \right. \right. \nonumber \\
&& \left. \left. -\varepsilon(n) 
\cos(M_W \pi R) \sqrt{\cos(\varphi_{n} + \alpha_{n})} \sin \varphi_{n}  
\right) \cos(\sqrt{m_{n\pm}^2-M^2}y)
\right] \nonumber \\
&=& 
(-1)^n
\left[ 
\cos(M_W y) \sqrt{\cos(\varphi_{n} + \alpha_{n})}  
\cos(\sqrt{m_{n}^2-M^2}y-\alpha_{n}) \right. \nonumber \\
&& \left. 
+ \varepsilon(n) \sin(M_W y) \sqrt{\cos(\varphi_{n} - \alpha_{n})} \sin(\sqrt{m_{n}^2-M^2}y)
\right] \nonumber \\
&=& (-1)^n [F_{-M, M_W}^{(n)}(y)/C^{(n)}]_{{\rm real}} \nonumber \\
&=& (-1)^n [f_{d_R}^{(n)}(y)/C^{(n)}]_{{\rm real}}
\eea
where 
we used the relations
\bea
\sin(M_W \pi R) \cos \alpha_{n} &=& \varepsilon(n) (-1)^n \sin \varphi_{n}, \\
\cos(M_W \pi R) \sin \varphi_{n} &=& 
\varepsilon(n) \sin(M_W \pi R) \sqrt{\cos(\varphi_{n} \pm \alpha_{n})} 
\sqrt{\cos(\varphi_{n} \mp \alpha_{n})} 
\eea
which can be derived from
\bea
\sin^2 \varphi_n = \sin^2(M_W \pi R) \cos^2 \alpha_n 
= \tan^2(M_W \pi R) \cos(\varphi_n+\alpha_n) \cos(\varphi_n-\alpha_n). 
\eea
A similar relation for the imaginary part can be verified. 
Thus, we obtain 
\bea
&&\int_{-\pi R}^{\pi R} dy \cos \left(\frac{m}{R}y \right) f_{d_R}^{(n)}(y)^* f_{d_R}^{(0)}(y) \nonumber \\
&=& 2 \int_0^{\pi R} dy \cos \left(\frac{m}{R}y \right) 
\left[
f_{d_R}^{(n)}(y)_{{\rm real}} f_{d_R}^{(0)}(y)_{{\rm real}} + f_{d_R}^{(n)}(y)_{{\rm imaginary}} f_{d_R}^{(0)}(y)_{{\rm imaginary}} 
\right] \nonumber \\
&=& (-1)^n 2\int^{\pi R}_{0} dy (-1)^m \cos \left(\frac{m}{R}(y-\pi R) \right) \times \nonumber \\
&& \left[
f_{d_L}^{(n)}(y-\pi R)_{{\rm real}} f_{d_L}^{(0)}(y-\pi R)_{{\rm real}} 
+ f_{d_L}^{(n)}(y-\pi R)_{{\rm imaginary}} f_{d_L}^{(0)}(y-\pi R)_{{\rm imaginary}} 
\right] \nonumber \\
&=& (-1)^{n+m} \int_{-\pi R}^{\pi R} dy \cos \left(\frac{m}{R}y \right) f_{d_L}^{(n)}(y)^* f_{d_L}^{(0)}(y),
\eea
which concludes the fact $I_c^{nm}(-M, M_W; -M, M_W) = (-1)^{n+m} I_c^{nm}(M, M_W; M, M_W)$. 

\section{The ``equivalence theorem"}

In this appendix,  
we discuss the relation between the vertex functions of $D$-dimensional gauge boson and scalar $(A_{\mu}, A_{y})$, 
which are the partners of the Higgs-like mechanism in the sector of non-zero K-K modes 
or form a multiplet of quantum mechanical SUSY. 
The relation is what we naturally expect from the fact 
that the interaction of the longitudinal component of the massive gauge boson is equivalent to 
that of the would-be Nambu-Goldstone boson. 
As the typical example, we will derive a relation between 
$I_{c}^{(nm)}$ and $I_{s}^{(nm)}$ defined in (\ref{3.6}) and (\ref{3.13}). 

A key ingredient is the fact that right- and left-handed Weyl fermions form a multiplet of quantum mechanical SUSY, 
whose transformation is given as (see (\ref{2.35}) and (\ref{2.36})) 
\bea 
&& i(\partial_{y} -iM_{W} -M \epsilon (y)) f_{dR}^{(n)} (y) = m_{n} f_{dL}^{(n)} (y), \\  
&& i(\partial_{y} -iM_{W} +M \epsilon (y)) f_{dL}^{(n)} (y) = m_{n} f_{dR}^{(n)} (y).  
\eea
By use of these relations we can verify (for $m \geq 1$)
\bea 
&& I_{s}^{(nm)}(-M, M_{W}; M, M_{W}) 
= \frac{1}{\sqrt{\pi R}} \int_{-\pi R}^{\pi R} dy \sin (\frac{m}{R}y) f_{dR}^{(n)}(y)^{\ast} 
f_{dL}^{(0)}(y) \nonumber \\ 
&=& \frac{1}{\sqrt{\pi R}} \int_{-\pi R}^{\pi R} dy \frac{1}{(\frac{m}{R})} \cos (\frac{m}{R}y) 
[(\partial_{y} f_{dR}^{(n)}(y)^{\ast})f_{dL}^{(0)}(y) + f_{dR}^{(n)}(y)^{\ast}(\partial_{y} f_{dL}^{(0)}(y))] \nonumber \\ 
&=& \frac{1}{\sqrt{\pi R}} \int_{-\pi R}^{\pi R} dy \frac{1}{(\frac{m}{R})} \cos (\frac{m}{R}y) 
[((\partial_{y} +iM_{W} - M \epsilon (y))f_{dR}^{(n)}(y)^{\ast})f_{dL}^{(0)}(y) \nonumber \\ 
&& \hspace*{10mm} + f_{dR}^{(n)}(y)^{\ast}((\partial_{y} -iM_{W} + M \epsilon (y))f_{dL}^{(0)}(y))] \nonumber \\ 
&=& \frac{i}{\sqrt{\pi R}} \int_{-\pi R}^{\pi R} dy \frac{1}{(\frac{m}{R})} \cos (\frac{m}{R}y) 
[m_{n} f_{dL}^{(n)}(y)^{\ast}f_{dL}^{(0)}(y) - m_{d} f_{dR}^{(n)}(y)^{\ast}f_{dR}^{(0)}(y)] \nonumber \\ 
&=& i \frac{1}{(\frac{m}{R})} [m_{n} - (-1)^{m+n} m_{d}] I_{c}^{(nm)} (M, M_{W}; M, M_{W}),   
\eea
where the step to move to the second line is due to a partial integral. 
This relation immediately leads to the relation between the vertex functions,  
\bea 
V_{\gamma_y}^{nm} = i \frac{m_{n} - (-1)^{m+n} m_{d}}{(\frac{m}{R})} V_{\gamma_{\mu}}^{nm} \ \ (m \geq 1). 
\eea

\end{appendix}

\end{document}